\pdfoutput=1
\documentclass[10pt]{article}

\usepackage[utf8]{inputenc}
\usepackage{cite}
\usepackage{hyperref}
\hypersetup{colorlinks=true,linkcolor=blue2,citecolor=red2,urlcolor=green2,pdfencoding=auto,linktocpage}
\usepackage{amsmath,amssymb,amsbsy,amstext,amsthm,simplewick,amsfonts}
\usepackage{mathrsfs}
\usepackage{graphicx}
\usepackage{wrapfig}
\usepackage{upgreek}
\usepackage{bm} 
\usepackage{framed}
\usepackage{bbm}
\usepackage{textcomp}
\usepackage{adjustbox}
\usepackage{makecell}
\usepackage{tcolorbox}
\usepackage{empheq}
\usepackage[normalem]{ulem}
\usepackage{enumitem}
\usepackage{braket} 
\usepackage{array}
\usepackage{dsfont}
\usepackage{physics}
\usepackage{ulem}
\usepackage{xcolor}
\usepackage{caption}
\usepackage{subcaption}
\usepackage{tocloft}
\usepackage{tikz}
\usetikzlibrary{positioning,arrows.meta}
\usepackage{multirow}

\usepackage[framemethod=default]{mdframed}

\newmdenv[backgroundcolor=gray!15,%
skipabove=5pt,%
skipbelow=5pt,%
leftmargin=2pt,%
rightmargin=2pt,%
innertopmargin=-6pt,%
innerbottommargin=5pt,%
innerleftmargin=5pt,%
innerrightmargin=5pt,%
splittopskip=0pt,%
splitbottomskip=0pt,%
linewidth=0pt,%
nobreak=true]%
{keyeqn}

\graphicspath{{Fig/}}

\definecolor{red2}{RGB}{214, 39, 40}
\definecolor{green2}{RGB}{0,170,0}
\definecolor{blue2}{RGB}{0,100,200}
\definecolor{magenta2}{RGB}{191,64,191}
\definecolor{purple2}{RGB}{112,48,160}
\definecolor{orange2}{RGB}{255,192,0}

\def\d{\mathrm{d}}

\def\bfk{\mathbf{k}}
\def\bfp{\mathbf{p}}
\def\bfq{\mathbf{q}}
\def\bfx{\mathbf{x}}

\def\Re{\mathrm{Re}\,}
\def\Im{\mathrm{Im}\,}

\newcommand{\mcO}{{\mathcal{O}}}
\newcommand{\mcM}{{\mathcal{M}}}
\newcommand{\mcD}{{\mathcal{D}}}
\newcommand{\deltabr}{{[\delta]}}
\newcommand{\dbar}{{\tilde\dd}}
\newcommand{\deltabar}{{\tilde\delta}}

\newcommand{\EFT}{\text{EFT}}

\renewcommand{\O}{\mathcal{O}}
\newcommand{\ceq}{\coloneqq}
\newcommand{\eqc}{\eqqcolon}

\numberwithin{equation}{section}

\allowdisplaybreaks[1]
\newmuskip\pFqskip
\mathchardef\pFcomma=\mathcode`, 

\newcommand*\pFq[5]{%
  \begingroup
  \begingroup\lccode`~=`,
    \lowercase{\endgroup\def~}{\pFcomma\mkern\pFqskip}%
  \mathcode`,=\string"8000
  {}_{#1}F_{#2}\biggl[\genfrac..{0pt}{}{#3}{#4};#5\biggr]%
  \endgroup
}

\begin{document}

\begin{titlepage}
	\setcounter{page}{1} \baselineskip=15.5pt 
	\thispagestyle{empty}

     \begin{center}
		{\fontsize{22}{22}\centering {\bf{On the Asymptotic Nature \\[0.3cm] of Cosmological Effective Theories}}}
	\end{center}

 
	\vskip 18pt
	\begin{center}
		\noindent
		{\fontsize{12}{18}\selectfont Carlos Duaso Pueyo\footnote{\tt carlos.duasopueyo@sns.it}$^{,ab}$, Harry Goodhew\footnote{\tt goodhewhf@ntu.edu.tw}$^{,cd}$ Ciaran McCulloch\footnote{\tt cam235@cam.ac.uk}$^{,e}$ and Enrico Pajer\footnote{\tt enrico.pajer@gmail.com}$^{,e}$}
	\end{center}
	
	\begin{center}
		\vskip 12pt
        $^a$ \textit{Scuola Normale Superiore, Piazza dei Cavalieri 7, 56126 Pisa, Italy} \\
        $^b$ \textit{INFN, Sezione di Pisa, Largo Bruno Pontecorvo 3, 56127 Pisa, Italy}\\
        $^c$ \textit{Leung Center for Cosmology and Particle Astrophysics, Taipei 10617, Taiwan}\\
        $^d$ \textit{Center for Theoretical Physics, National Taiwan University, Taipei 10617, Taiwan} \\
        $^e$ \textit{Department of Applied Mathematics and Theoretical Physics, University of Cambridge,\\Wilberforce Road, Cambridge, CB3 0WA, UK} 
	\end{center}
	
	\vskip 50pt
	\noindent\rule{\textwidth}{0.4pt}
	\noindent \textbf{Abstract} Much of our intuition about Effective Field Theories (EFTs) stems from their formulation in flat spacetime, yet EFTs have become indispensable tools in cosmology, where time-dependent backgrounds are the norm. In this work, we demonstrate that in spacetimes undergoing significant expansion—such as accelerated FLRW and de Sitter backgrounds—the contributions of operators with mass dimension $\Delta$ to physical observables grow factorially with $\Delta$ at fixed couplings. This behavior stands in stark contrast to static flat spacetime. As a result, the cosmological EFT expansion is generally asymptotic rather than convergent, even at tree level.
    
    To illustrate this phenomenon, we analyze simple toy models involving a massless or conformally coupled scalar field interacting with a heavy scalar with zero or infinite sound speed. We demonstrate that meaningful EFT predictions can still be extracted via appropriate resummation techniques, performed in both Fourier and Mellin-momentum space. In the infinite sound speed limit, where the heavy field is effectively non-dynamical, the resummed EFT reproduces the exact result of the full theory. In other cases, the EFT captures only the local part of the dynamics, omitting nonlocal terms, which are exponentially suppressed in the large-mass limit for the Bunch-Davies state. Our results provide detailed and quantitative expectations for more general cosmological EFTs.
	
	\noindent\rule{\textwidth}{0.4pt}
	
	
\end{titlepage} 


\newpage
\setcounter{page}{2}
{
	\tableofcontents
}
\setcounter{footnote}{0} 



\section{Introduction}\label{IntroSect}

Our starting point is the observation that, in the presence of a time dependent background, such as in cosmological spacetimes, operators of very high dimension in an Effective Field Theory (EFT), which are expected to be very irrelevant, at face value give very large contributions to low energy observables. We show this in detail for scalar EFTs in de Sitter spacetime, envisaging applications to cosmology. To be more precise, we begin with an elementary review of EFTs in time-independent flat spacetime and contrast this situation with a simple explicit calculation in a time-dependent cosmological background. While time dependence is not unique to curved spacetime, it is not often discussed in the context of flat-space scattering amplitudes. Conversely, time dependence is a universal defining feature of cosmological calculations and should be systematically accounted for when employing EFT principles.

\paragraph{Static flat spacetime} When a system with a characteristic energy scale $M$ is studied at a much lower energy scale $E$, a remarkable simplification occurs. EFTs provide a general and successful paradigm to leverage the separation of scales $E\ll M$ and deliver a simpler description of the system that focuses only on the relevant degrees of freedom. The construction of an EFT requires schematically three steps: (i) Identify the relevant degrees of freedom and symmetries; (ii) Write down the infinite number of effective interactions built out of the degrees of freedom and compatible with the symmetries; (iii) Establish a power counting to estimate the contribution from each interaction and use it to truncate the theory to a finite number of interactions. For the theory to be predictive a fourth step is necessary in which one fixes the values of the unknown low-energy constants either by fitting to experiments (bottom up) or by matching to some high-energy completion of the EFT (top down). For concreteness and following \cite{Polchinski:1992ed}, we will now briefly review the case of a time-independent system in flat spacetime. Let's consider an operator $\mathcal{O}_\Delta$ of dimension $\Delta$ and a single-scale power counting in which this appears in the action as
\begin{align}
 S\supset \int \dd^4x g \frac{\O_\Delta}{M^{\Delta-4}}\,,
\end{align}
where $g$ is a dimensionless coupling that is expected to be of order one if naturalness applies. When probing the theory at energy $E$ we expect $d^4x\sim E^{-4}$ and $\O_\Delta\sim E^\Delta$, so that the contribution of this operator to observable is expected to be of order 
\begin{align}\label{expect}
    \text{Observable}\sim g \left(\frac{E}{M} \right)^{\Delta-4}+\dots
\end{align}
where the dots contain higher order contributions in $g$. Irrelevant operators with $\Delta-4>0$ give smaller and smaller contributions to low energy observables. An elementary toy model confirming the above expectation is the 2-to-2 scattering amplitude of a light field $\phi$ mediated by the tree-level exchange of a field $\chi$ with mass $M$. For a coupling $(\lambda/2) \phi^2 \chi$ the tree-level $s$-channel contribution is simply\footnote{It is straightforward to include the other channels $(T-M^2)^{-1}$ and $(U-M^2)^{-1}$. For the physical process $12\to 34$ note that $S>4m^2$---where $m$ is the mass of the light field---and $T,U<0$. From $S+T+U=4m^2$ it follows that $|T|,|U|< S $.}
\begin{align}\label{pred}
    A_{4s}=\frac{\lambda^2}{S-M^2}\,,
\end{align}
where $S=-(p_1+p_2)^2$ is the first Mandelstam variable. When probing the theory at an energy $E=\sqrt{S}$ much smaller than its characteristic scale $M$, the simplification is that we don't need to include the field $\chi$ in the theory because we can never reach the pole $s=M^2$. Instead, we can approximate the contribution of this pole with higher derivative quartic interactions of the schematic form $\O_{\Delta}\sim (\lambda/M)^2\phi^2 (\Box/M^2)^n \phi^2$, where $\Delta = 2n-4$ leading to
\begin{align}
    A_{4s}=\frac{\lambda^2}{M^2}\sum_{n=0}^\infty \left(\frac{S}{M^2}\right)^n\,. \label{eq:A4EFT}
\end{align}
Not only can the EFT with just $\phi$ reproduce the tree-level prediction of the full theory~\eqref{pred} to arbitrary accuracy (at order $\lambda^2$), but also a finite number of higher dimension operators is needed for any finite precision. In other words, the contribution of $\O_{\Delta}$ to $A_{4s}$ scales as $(E/M)^{\Delta-4}$, as predicted by~\eqref{pred}, and so becomes arbitrarily small for fixed $\Delta>4$ as $(E/M) \to 0$ or for fixed $E/M<1$ as $\Delta \to \infty$. In yet other words, the Taylor expansion of $A_{4s}$ around $S=0$ has a finite radius of convergence $|S|=M^2$.\\

\paragraph{Time-dependent background} The reason why we have dragged the time-starved reader through such an elementary discussion is to point out that in a time-dependent background things can work out very differently. Indeed, let's consider the toy model of a massless scalar field in de Sitter spacetime. Just like in the example above we imagine the massless scalar field is coupled to a massive scalar $\chi$, but to avoid having to concern ourselves with late-time divergences, we assume $\phi$ has a shift symmetry and consider the coupling $(\lambda /2) \chi \dot\phi^2$. The exact result for the late-time four-point correlator of $\phi$ from the exchange of $\chi$ is known \cite{Arkani-Hamed:2018kmz}, but we want to focus on the EFT where $\chi$ has been integrated out. This EFT should contain the infinitely many higher dimension operators. For concreteness, let's focus on the following infinite family of higher-dimension operators for which the algebra is exceptionally simple (see Section~\ref{sec2:EFT} for more details)
\begin{align} \label{eq:EFTops}
    \mathcal{L}_\text{EFT}\supset -\frac{\lambda^2}{4M^2} {\phi'}^2  \left(\frac{\nabla^2}{a^2M^2}\right)^n\left({\phi'}
    ^2\right).
\end{align}
It is straightforward to compute the four-point function induced by one such contact operator in the in-in (or in-out \cite{Donath:2024utn}) formalism: 
\begin{equation} 
\begin{aligned}
    B_{4s}^{(n)}&=2\Re \frac{2i\lambda^2s^{2n}}{M^{2(n+1)}}\int \frac{\dd\eta}{(\eta H)^4} (H\eta)^{2n} \left[\prod_a^4(-\eta H \partial_\eta) G_{+}(\eta,k_a)\right] \label{uno}\\
    &= \frac{\lambda^2}{M^2} \frac{H^8}{4 k_1k_2k_3k_4 k_T^5}\left(\frac{Hs}{Mk_T}\right)^{2n}\Gamma(\Delta-3) \, ,
\end{aligned}
\end{equation}
where $s=|\bfk_1+\bfk_2|$ and $\Delta=8+2n$ is the dimension of the operator. This result is largely familiar: the overall scaling $B_{4s}\sim k^{-9}$ is enforced by de Sitter dilations, the only pole is the total-energy pole $k_T\ceq \sum_a |\bfk_a|=0$, and the order of the pole is $1+(\Delta-4)$ as expected on general grounds \cite{Pajer:2020wxk}. If we strip off the kinematics\footnote{This can be done more precisely considering the field-theoretic wavefunction $\Psi$, where integrals over the Fourier modes $\bfk_a$ ensure that $\log \Psi$ scales as $k^0$.}, we find the suppression factor $(H/M)^\Delta$ for large $\Delta$. This confirms the intuition\footnote{The reason why we don't find the precise exponent $\Delta-4$ in~\eqref{expect} is that time and space derivatives don't scale in the same way as this is not a single-scale power counting. We shouldn't get distracted by this technical detail.} that cosmological correlators of massless fields probe the theory at an energy of order Hubble. What strongly differs from flat spacetime and prompted this investigation is the factorial enhancement $\Gamma(5+2n)$ of large dimension operators. This enhancement is forthright disturbing: no matter how small $H/M$ is, there are always infinitely many very-high-dimension operators whose contribution to the observable correlator $B_{4s}$ is much larger than that of any lower order ones. Naively, these very large contributions also seem to be incompatible with perturbativity of the effective field theory \cite{DuasoPueyo:2024twm}. This simple example shows that in general the EFT expansion is not convergent. While this fact is not unique to curved spacetime and would occur also in time-dependent situations in flat spacetime, it is a universally occurring feature of cosmological theories and calculations. We hope that our detailed discussion of several cosmological toy models in this work will help setting the correct expectations for more general cosmological EFTs.

\paragraph{Summary of results} As the reader might have guessed, the sum of EFT contributions~\eqref{uno} is asymptotic and can be resummed with a variety of mathematical techniques, the better known of which is due to Borel. The goal of this paper is to study how the EFT contributions can be made sense of via resummation techniques and to what extent the resummation agrees with the (partial) high-energy completion of the EFT. To do this, we focus on two de Sitter toy models which can capture the main qualitative features while avoiding the technical difficulty of a fully realistic theory. Here is a summary of our main results:
\begin{itemize}
    \item In Section~\ref{sec:NonDynField} we study the EFT expansion for a non-dynamical massive field coupled to a light field in de Sitter. We show that Mittag-Leffler summation of the EFT contributions to the wavefunction correctly recovers the UV result.
    \item In Section~\ref{sec:zero cs} we switch to the case of a massive field with zero sound speed coupled to a light field in de Sitter. In this case the resummation of the EFT contributions misses terms of the UV result. Such terms are exponentially small in $M/H$ if the massive field is in the Bunch-Davies state, but can be larger for an excited initial state. We explicitly show how the missed terms can be accounted for in the EFT via ad hoc boundary contributions.
    \item The resummed result is compatible with UV unitarity in the non-dynamical field case, as pointed out in Section~\ref{sec:nondynCOT}. In contrast, Section~\ref{sec:zerocsCOT} shows that for the zero sound speed case the EFT resummation does not satisfy the UV unitarity condition—it misses the information about the discontinuity in the full theory.
    \item In Section~\ref{sec:Mellin} we study the zero sound speed EFT expansion in Mellin space and show that it actually converges, offering an alternative summation method. The Mellin resummation procedure presents an ambiguity corresponding to different possible boundary conditions of the massive field, capturing the fact that such information is not accessible from the EFT. 
\end{itemize}

\paragraph{Comments} Before getting into the details of our study, several general remarks are in order. 

First, the asymptotic nature of the perturbative expansion in quantum field theory is of course very well known \cite{Dyson:1952tj,Lipatov:1976ny,tHooft:1977xjm,Bender:1969si}. The non-convergence of the perturbative expansion comes about because the sum over the different order solutions of the equations of motion is exchanged in perturbative calculation with the average implemented by the path integral or stochastic initial conditions. The average ranges over values of the fields that extend beyond the radius of convergence of the perturbative solution and this results in an ever increasing mistake to higher orders in perturbation theory. The issue we are studying in this paper is not unrelated to but quite distinct from this. Here we point out that integrating out a field \textit{already at tree level}, generates a infinite number of higher dimension operators whose contribution to low energy observables grows factorially in the mass dimension, a phenomenon that would not occur in the time-independent background of Minkowski spacetime. 

The second comment is that even though all of our detailed examples concern higher \textit{derivative} (quartic) operators in de Sitter spacetime, the phenomenon we are describing is much more general and would occur for general higher \textit{dimension} operators, with arbitrary field multiplicity, and for general accelerating\footnote{Similar issues could also arise in decelerating cosmologies. Here we stick to accelerated cosmologies because in that case it is clear how to consistently impose a Bunch-Davies initial state in the infinite past.} FLRW spacetimes. For example, consider an interaction of the schematic form $\partial^m \phi^n \Lambda^{4-n-m}$ in de Sitter. Just like in~\eqref{uno} we get a power of $\eta$ for each derivative, but also an additional power of $\eta$ for each field since, in the early-time limit the mode functions of a field of arbitrary mass asymptote $\eta e^{ik\eta}$. Upon performing the time integral this in turn leads to the factorial growth $\Gamma(\Delta-3)$ where $\Delta=n+m$ is the dimension of the operator. Another example is a conformally coupled field with interaction of the schematic form $\partial^m \varphi^n \Lambda^{4-n-m}$ in an FLRW spacetime with scale factor $a=(\eta_*/\eta)^\alpha$, where $\eta$ is conformal time. This spacetime experiences accelerated expansion, i.e. $\ddot a>0$, iff $\alpha>0$, with $\alpha (\alpha-1)\geq 0$ representing the null-energy condition (NEC). The corresponding contact correlator is
\begin{align}
   B_n &\sim 2\Re i a(\eta_0)^{-n} \int_{-\infty}^{\eta_0} \dd\eta \, a^{4-\Delta} \frac{k^m}{\prod_{a=1}^n k_a}e^{ik_T \eta} \\
   &\propto  \left(\frac{\eta_0}{\eta_*}\right)^{\alpha n} \frac{k^m \Gamma[\alpha(\Delta-4)+1]}{k_T \prod_{a=1}^n k_a} \Re \left[ \left(\frac{i}{k_T \eta_*} \right)^{\alpha(\Delta-4)} \left( 1 + i k_T \eta_0 + O\left( (k_T \eta_0)^2\ \right)\right) \right]\,,
\end{align}
where $k_T=\sum_a^n k_a$. The factorial enhancement is now $\Gamma[\alpha(\Delta-4)+1]$, which for NEC-abiding accelerated cosmologies grows even faster than factorially in $\Delta-4$. 

A third comment is that there is a very direct way to see that a cosmological correlator has a factorial enhancement relative to the corresponding flat space amplitude. Indeed, the correlators $B_n$ of $n$ massless fields in de Sitter with a Bunch-Davies initial state display a characteristic pole when the so-called total energy vanishes $k_T=0$, which can be reached only for complex kinematics, whose residue is proportional to (the UV-limit of) the associated flat space $n$-particle amplitude $A_n$ \cite{Raju:2012zr,Maldacena:2011nz}. The precise relation including the overal normalization is \cite{Goodhew:2020hob}
\begin{equation}\label{finalBtoA}
\lim_{k_T\rightarrow 0} B_n=2\,(-1)^{n} H^{2I+2n+\sum\limits_\alpha^V(N_\alpha-4)} \Gamma[p]\times \Re\left\lbrace  \frac{iA_n'}{(-ik_T)^{p} }\prod_a^n\frac{i+k_a\eta_0}{2k_a^2}\right\rbrace,
\end{equation}
where $I$ is the number of internal lines, $p=1+\sum_\alpha (\Delta_\alpha-4)$ with the sum running over all $V$ vertices in the associated diagram, which have mass dimension $\Delta_\alpha$ and multiplicity $N_\alpha$. This relation tells us that $B_n$ is enhanced as compared to $A_n$ by $(p-1)!$, which becomes very large for operators of very large dimension. Even when the flat-space EFT expansion is convergent for $A_n$, the corresponding cosmological EFT expansion for $B_n$ is typically not. 


\paragraph{Relation to previous work} We are not the first ones to notice the asymptotic behavior of the derivative expansion in cosmological spacetimes and here we comment on a series of related results and investigations in the literature.

Particle production in time-dependent backgrounds is a well-studied physical effect. Characterizing it, for example by computing Bogoliubov coefficients and particle number densities, requires the study of the limiting behavior of mode functions at early and late times. Due to the non-perturbative nature of particle production these mode functions manifest the Stokes phenomenon, so that their description as a perturbative expansion is asymptotic (see for instance~\cite{Hashiba:2021npn,Sou:2021juh} and references therein). In this work we point out that the derivative expansion resulting from integrating out a massive field in cosmology is similarly asymptotic.

Effective field theories in time-dependent settings were also explored in the past. For example, in \cite{Collins:2012nq} the authors use the in-in (Schwinger-Keldysh) formalism and analyze a model comprising interacting heavy and light scalar fields in flat spacetime, both initialized in their free vacuum states at a \textit{finite} initial time. They find that the EFT for the light scalar contains terms that depend on the initial conditions, a phenomenon for which they use the arguable expression ``non-locality in time". Three crucial differences with our study are that we focus on de Sitter spacetime, that we impose an initial Bunch-Davies state \cite{Bunch:1978yq} in the infinite past of the Poincar\'e patch (which can be reached in a finite \textit{proper} time), as opposed to at some finite time, and that we point out the asymptotic nature of the derivative expansion. 

The divergent behavior of a particular time-derivative expansion was pointed out in~\cite{Jazayeri:2023kji}, where a model with a heavy, luminal vector field coupled to a massless, slowly propagating scalar in de Sitter was studied. In this peculiar setup 
the EFT that results from integrating out the former is organized as an expansion in (only) time derivatives of the latter, and is non-local in space. The authors mention that such a series is formally divergent, since the time derivatives' contribution grows factorially, and that this reflects the failure to capture the on-shell production of the heavy field. Here we show that this phenomenon is more general, and is not necessarily tied to time derivatives. As a matter of fact, the model that we study in Section~\ref{sec:NonDynField} corresponds to keeping only the zeroth-order operator in the time-derivative expansion of~\cite{Jazayeri:2023kji}, which has $(\partial_t)^0$, and the asymptotic behavior is observed too.

Finally, the question of the convergence of the EFT derivative expansion in flat space was studied in~\cite{Morris:1997km,Morris:1999ba,Litim:2001dt}, in the context of attempting to solve the RG flow equations approximately by including contributions from operators in the effective action up to some order in derivatives. Differently from our case, the issue there is that these contributions are evaluated in the flow equations at a scale close to the radius of convergence of the derivative expansion, so special care is needed to avoid divergent behavior. It would be interesting to see what the asymptotic behavior described in this paper implies for these techniques, were they to be applied in curved spacetimes.

\paragraph{Notation and conventions} We work in 3+1 dimensions, using conformal coordinates on a flat slicing,
\begin{equation}
    ds^2 = a(\eta)^2 (-d\eta^2+d\mathbf{x}^2 ) \, ,
\end{equation}
with $a(\eta)=-1/H\eta$ for de Sitter spacetime. Derivatives with respect to $\eta$ are sometimes indicated with a prime. The field theoretic wavefunction at the future boundary of de Sitter (conformal time $\eta=0$) is parameterized as
\begin{align}
    \Psi[\phi]=\exp\left[   -\sum_{n}^{\infty}\frac{1}{n!} \int_{\bfk_{1},\dots,\bfk_{n}}\,  (2\pi)^{3} \delta^{(3)} \left( \sum_{a}^{n} \bfk_{a} \right) \psi_{n}(\{\bfk\}) \, \phi(\bfk_{1})\dots \phi(\bfk_{n})\right]\,,
\end{align}
where $\psi_n$ are wavefunction coefficients. To simplify the notation we use
\begin{equation}
    \int_\bfk \ceq \int\frac{\dd^3 k}{(2\pi)^3} \, ,\qquad\text{and}\qquad \int_\bfx \ceq \int\dd^3 x \, .
\end{equation}
We define four-point correlators as
\begin{align}
    \langle \phi_{\mathbf{k}_1}\phi_{\mathbf{k}_2}\phi_{\mathbf{k}_3}\phi_{\mathbf{k}_4} \rangle_s = (2\pi)^3 \, \delta\left(\sum_{a=1}^4\bfk_a\right) \, B_{4s} (\bfk_1,\bfk_2,\bfk_3,\bfk_4) \,,
\end{align}
where the label $s$ refers to the $s$-channel. We use that label not only for quantities coming from exchange diagrams but also from contact diagrams, meaning that we only write the contribution that has the same permutation symmetry as an $s$-channel exchange diagram (see e.g.~\eqref{eq:A4EFT} and~\eqref{uno}). We work in momentum space, using $k_a\ceq |\bfk_a|$ and $k_{ab} \ceq k_a+k_b$. In four-point wavefunction coefficients and correlators we also use
\begin{equation}
    k_T \ceq \sum_{a=1}^n k_a \, , \qquad s\ceq |\bfk_1+\bfk_2| = |\bfk_3+\bfk_4| \, , \qquad e_4 \ceq k_1k_2k_3k_4 \, .
\end{equation}
The conventions used in Mellin space are explained in Appendix~\ref{app:Mellin}. 


\section{Non-dynamical massive field} \label{sec:NonDynField}

We start with the simplest example, a massive field $\chi$ whose equations of motion contain no time derivatives. We will couple this to a shift symmetric scalar field $\phi$, therefore, the Lagrangian for the theory is
\begin{align} \label{eq:NonDynField}
    \mathcal{L}= - \frac{1}{2} a^4 (\partial_{\mu}\phi)^2 - \frac{1}{2}a^2\partial_i\chi\partial^i\chi-\frac{1}{2}M^2 a^4\chi^2+\frac{\lambda}{2} a^2{\phi'}^2\chi.
\end{align}
This field can be integrated out just like any other, with the advantage that the resulting EFT will contain only spatial derivatives. This EFT, as already mentioned, captures (at leading order in time derivatives) the dynamics of the low-speed cosmological collider~\cite{Jazayeri:2022kjy,Jazayeri:2023kji}. One can think of this action as arising from the limit in which the speed of sound of $\chi$ is very large (as compared to that of $\phi$) and so one can neglect the time-derivative part of the $\chi$ kinetic term. 


\subsection{Calculation in the full theory}

The equations of motion for $\chi$ are
\begin{align}
    (-a^2\nabla^2 +a^4M^2)\chi=-\frac{\lambda}{2} a^2{\phi'}^2.
\end{align}
If we demand that this field vanished in the infinite past, so that we are in a vacuum state, then the mode functions are zero at all times by the non-dynamical nature of the field. However, there is still a contribution from the Green's function which satisfies the equation
\begin{align} \label{eq:nondynGreen}
    (-a^2\nabla^2 +a^4M^2)G_\chi(x,x')=-i\delta^4(x-x')\,,
\end{align}
subject to the boundary condition that the field vanishes in the infinite past, on the future boundary, and at spatial infinity. It will be convenient to perform calculations in momentum space where the propagator is
\begin{align}
    G_\chi(k;\eta,\eta')=\frac{-iH^4\eta^4}{k^2H^2\eta^2+M^2}\delta(\eta-\eta')\,.
\end{align}
This form of the propagator is suggestive of the fact that $\chi$ mediates instantaneous interactions at distance, as expected for a field with a very high speed of sound. The $s$-channel quartic wavefunction coefficient $\psi_{4s}$ in the UV theory is therefore, 
\begin{align}
    \psi_{4s}&=(-i\lambda)^2\int \dd\eta \dd\eta' a^2 a'^2\big(K^{\phi}_{k_1}(\eta)\big)'\big(K^{\phi}_{k_2}(\eta)\big)'\big(K^{\phi}_{k_3}(\eta')\big)'\big(K^{\phi}_{k_4}(\eta')\big)'\frac{-iH^4\eta^4}{s^2H^2\eta^2+M^2}\delta(\eta-\eta') \label{eq:nondynexactcalc} \\
    &=i\frac{\lambda^2k_1^2k_2^2k_3^2k_4^2}{M^2}\int \dd\eta \frac{\eta^4e^{ik_T\eta}}{1+k_T^2z^2\eta^2} =-\frac{\lambda^2k_1^2k_2^2k_3^2k_4^2}{M^2k_T^5} \, \frac{2z(1+2z^2)+e^{-1/z} \, {\rm E}_1(-1/z) - e^{1/z} \, {\rm E}_1(1/z)}{2z^5} \, . \nonumber
\end{align}
where $s\ceq |\bfk_1+\bfk_2|$, ${\rm E}_1 (w)$ is the exponential integral, and $K^{\phi}$ is the usual bulk-boundary propagator for the massless field $\phi$,
\begin{align}
    K_k^{\phi}(\eta)=(1-ik\eta)e^{ik\eta}\, .
\end{align}
We have also introduced 
\begin{align}
    z \ceq \frac{Hs}{Mk_T} \, ,
\end{align}
for notational simplicity, and have assumed that $k_T$ has a small negative imaginary part to recover the Bunch-Davies vacuum in the infinite past. 
Since the bulk-boundary propagator of $\chi$ vanishes, the cubic wavefunction coefficient $\psi_{\phi\phi\chi}$ for this theory also vanishes. Hence, the 4-point function of $\phi$ is simply proportional to the real part of $\psi_{4s}$\footnote{Notice that ${\rm E}_1(1/z)$ is real for $z\in\mathbb{R}_+$, but ${\rm E}_1(-1/z)$ is complex. Its imaginary part is sensitive to the $i\epsilon$ prescription that we implement on $k_T$ to make the time integral~\eqref{eq:nondynexactcalc} converge, which tells us that $k_T$ ($z$) approaches the positive real line from below (above). We have then ${\rm Im} \, [{\rm E}_1(-1/z)]=-\pi$. This imaginary part appears in the wavefunction but not in the correlator.},
\begin{align}
    B_{4s}= \frac{2\,{\rm Re} \,\psi_{4s}}{\prod_{a=1}^4 2\,{\rm Re} \,\psi_2^{\phi} (k_a)} =-\frac{\lambda^2 H^8}{16\, k_1k_2k_3k_4M^2k_T^5}\frac{2z(1+2z^2) + e^{-1/z}\,{\rm Re} \left[{\rm E}_1(-1/z)\right]-e^{1/z}\,{\rm E}_1(1/z)}{z^5} \, , \label{eq:B4nondyn}
\end{align}
where we have used that the two-point wavefunction coefficient for a field with bulk-to-boundary propagator $K_k(\eta)$ is
\begin{equation} \label{eq:psi2grformula}
    \psi_2 = -i \lim_{\eta\to\eta_0} a(\eta)^2 K_k(\eta) \, \partial_{\eta} K_k(\eta) \, .
\end{equation}


\subsection{Calculation in the EFT and resummation}\label{sec2:EFT}

The solution to the equations of motion for $\chi$ can also be written in terms of an infinite sum of local derivative operators acting on the massless field
\begin{align}
    \chi=-\frac{\lambda}{2} a^2\frac{1}{-a^2\nabla^2 +a^4M^2}\left({\phi'}
    ^2\right)=-\frac{\lambda }{2a^2M^2}\sum_{n=0}^\infty \left(\frac{\nabla^2}{a^2M^2}\right)^n\left({\phi'}
    ^2\right)+\chi_{\text{CF}}.
\end{align}
We have made explicit the possible additional term $\chi_{\text{CF}}$ that we are free to add when solving this equation, i.e. when we are inverting the singular equation of motion operator. This arbitrariness is always present when integrating out a field and to make progress one must specify the boundary conditions for the problem at hand (see Section~\ref{sec:EFT2} for a brief summary or Section 4 of \cite{salcedo2023analytic} for an extensive discussion of this in the context of the wavefunction). In our case, the boundary condition that $\chi$ vanishes in the past fixes $\chi_{\text{CF}}=0$ for all times. Similarly, the infinite sum vanishes in the late time limit and so we can ignore all boundary terms. This aspect will be very different in the model we discuss in the next section. Therefore, we will be working with the following single-field EFT
\begin{align} \label{eq:nondynEFT}
    \mathcal{L}_\text{EFT}= - \frac{1}{2} a^4 (\partial_{\mu}\phi)^2 -\frac{\lambda^2}{4M^2} {\phi'}^2\sum_{n=0}^\infty \left(\frac{\nabla^2}{a^2M^2}\right)^n\left({\phi'}
    ^2\right).
\end{align}
Inserting this back into the Lagrangian gives the wavefunction coefficients at a given order in our effective field theory\footnote{Note that the minus sign from $\nabla^2$ cancels with a minus sign coming from the additional factors of $\eta$ in the integral.},
\begin{align}
    \psi_{4s}^{(n)}=\frac{i\lambda^2s^{2n}}{M^{2(n+1)}}\int \dd\eta \, (H\eta)^{2n} \prod_{a=1}^4 \big(K^{\phi}_{k_a} (\eta) \big)' = \lambda^2 \frac{k_1^2k_2^2k_3^2k_4^2}{M^2k_T^5}\left(\frac{Hs}{Mk_T}\right)^{2n}\Gamma(5+2n) \, ,
\end{align}
As anticipated in the introduction, due to the time-dependence of the background we find that these contributions grow factorially as $\Gamma(\Delta-3)$ where $\Delta=8+2n$ is the mass dimension of the relevant operator containing $2n$ spatial derivatives, 4 time derivatives and 4 fields. At face value this suggests that higher dimensional operators give a larger contributions to $\psi_{4s}^\text{EFT}$ than lower dimensional ones. Moreover, the series obtained by summing over $n$ diverges and the naive result in the EFT is infinite
\begin{align}
    \psi_{4s}^\text{EFT}\overset{??}{=}\sum_{n=0}^{\infty} \psi_{4s}^{(n)} =\infty\,.
\end{align}
Here we point out that we can still make sense of this sum through the process of resummation. As the divergence goes like $(2n)!$ we use Mittag-Leffler summation, a generalization of Borel resummation (see Appendix~\ref{app:ML} for additional details), 
\begin{equation} \label{eq:nondynresum}
\begin{aligned}
    \psi_{4s}^{\text{EFT}}&=\int_0^\infty \dd t e^{-t} \left( \sum_{n=0}^\infty \psi_{4s}^{(n)} \frac{t^{2n}}{(2n)!} \right)= \frac{\lambda^2k_1^2k_2^2k_3^2k_4^2}{M^2k_T^5} \int_0^{\infty} \dd t e^{-t}\left(\frac{12}{(tz+1)^5}-\frac{12}{(tz-1)^5}\right)\\&=-\frac{\lambda^2k_1^2k_2^2k_3^2k_4^2}{M^2k_T^5}\frac{2z(1+2z^2)+e^{-1/z}\,{\rm E}_1(-1/z) -e^{1/z}\,{\rm E}_1(1/z)}{2z^5} \, .
\end{aligned}
\end{equation}
The convergence of this integral requires that $\Im(z)\neq0$ which is ensured by our Bunch-Davies condition that introduces a small positive imaginary part to $z=\frac{Hs}{Mk_T}$. This result is identical to what we saw in the exact calculation. Indeed, we can understand the resummation as moving the sum inside the time integral and hence undoing the mistake of exchanging sum and integral when we integrated out $\chi$. To see this, consider closing the time integral~\eqref{eq:nondynexactcalc} in the upper right quadrant to move the integration from $i\infty$ to $0$ and then redefine $t=-ik_T\eta$,
\begin{align} \label{eq:nondyn_psi4equal}
    \psi_{4s}&=i\frac{\lambda^2k_1^2k_2^2k_3^2k_4^2}{M^2}\int_{-\infty}^0 \dd \eta \frac{\eta^4e^{ik_T\eta}}{1+k_T^2z^2\eta^2}=\frac{\lambda^2k_1^2k_2^2k_3^2k_4^2}{2M^2k_T^5}\int_0^\infty e^{-t}\left(\frac{t^4}{1-z t}+\frac{t^4}{1+zt}\right)\dd t\\&=\frac{\lambda^2k_1^2k_2^2k_3^2k_4^2}{2M^2k_T^5}\int_0^\infty (\partial_t^4 e^{-t})\left(\frac{t^4}{1-zt}+\frac{t^4}{zt+1}\right)\dd t=\frac{\lambda^2k_1^2k_2^2k_3^2k_4^2}{M^2k_T^5}\int_0^\infty e^{-t}\left(\frac{12}{(tz+1)^5}-\frac{12}{(tz-1)^5}\right) \dd t \nonumber \\&=\psi_{4s}^{\text{EFT}} \, . \nonumber
\end{align}
The penultimate equality follows from integrating by parts repeatedly, noting that all of the boundary terms this generates will vanish. This shows us why the EFT is asymptotic. When we attempt to bring the sum in the Taylor expansion of $(1+k_T^2z^2\eta^2)^{-1}$ around $z=0$ outside of the time integral we encounter the obstacle that we are integrating over values of $\eta$ that are outside the radius of convergence of the expansion. This exchange of limits therefore results in an asymptotic series. The divergence of the series can be cured through resummation to recover the correct result. 

In this very simple theory there is no complementary function\footnote{By ``complementary" function we mean the general solution of the homogeneous equation of motion.} term that we could add that is compatible with our vacuum initial condition and there are no three point interactions. It is therefore trivial to read off the correlator in this theory as
\begin{align}
    B_{4s}^{\text{EFT}} = \frac{2\,{\rm Re} \,\psi_{4s}^{\text{EFT}}}{\prod_{a=1}^4 2\,{\rm Re} \,\psi_2^{\phi} (k_a)} \, .
\end{align}
Which, by virtue of~\eqref{eq:nondyn_psi4equal}, is in exact agreement with the full, UV-complete, theory result~\eqref{eq:B4nondyn}. The same resummation procedure works for a theory with conformally coupled field $\varphi$ and coupling $\varphi^2\chi$, rather than $\phi$ and $\phi'^{\,2}\chi$.


\subsection{Relationship to the Cosmological Optical Theorem} \label{sec:nondynCOT}

Unitarity plus the choice of the Bunch-Davies initial state\footnote{Similar results for other initial states were found in \cite{Cespedes:2020xqq,Ghosh:2024aqd,Ghosh:2025pxn}} implies that four-point wavefunction coefficients must satisfy the cosmological optical theorem (COT)~\cite{Goodhew:2020hob,Melville:2021lst,Goodhew:2021oqg}, which for the full theory~\eqref{eq:NonDynField} reads\footnote{Appendix C of~\cite{Stefanyszyn:2023qov} pointed out that, in this theory, the non-analytic behavior of the propagator~\eqref{eq:nondynGreen} in the internal energy variable implied an anomaly of Hermitian analyticity for $\psi_{4s}$, this was also discussed in~\cite{Goodhew:2024eup} through the lens of local symmetries in the Lagrangian. These results have important implications for the no-go theorems on parity-odd correlators \cite{Liu:2019fag,Cabass:2022rhr,Thavanesan:2025kyc}. However, this is not a problem here because the COT does not require us to analytically continue the internal energy $s$, but only the external ones.}
\begin{equation} \label{eq:nondynCOT}
    \psi_{4s} (k_a,s) + [\psi_{4s} (-k_a^*,s)]^* = 0 \, ,
\end{equation}
where the real values of $k_a$ should be approached from the lower-half complex plane \cite{Goodhew:2020hob}. In the full expression for the tree-level COT the right hand side would consist of two factors of the discontinuity of the three-point coefficient $\psi_{\phi\phi\chi}$, but it vanishes in this theory because $\psi_{\phi\phi\chi}=0$. To check that the result~\eqref{eq:nondynexactcalc} satisfies the COT, it is useful to express ${\rm E}_1(w)$ in terms of the complementary exponential integral ${\rm Ein} (w)$, which is an entire function, using
\begin{equation}
    {\rm E}_1 (w) = {\rm Ein} (w) - \ln w - \gamma \, ,
\end{equation}
where $w$ is a complex variable, $\gamma$ is Euler's constant, and the principal branch of the logarithm is to be taken. Using this and taking into account the $i\epsilon$ prescription for $k_T$ that makes the time integral~\eqref{eq:nondynexactcalc} converge, we have
\begin{equation} \label{eq:nondyn_pis4_Ein}
    \psi_{4s} \propto 2z(1+2z^2) + e^{-1/z}\bigg({\rm Ein}(-1/z) - \ln (1/z) -i\pi -\gamma \bigg) - e^{1/z}\bigg( {\rm Ein}(1/z)-\ln(1/z)-\gamma \bigg) \, ,
\end{equation}
where we have omitted an overall factor that does not play any role in the following. When substituting this four-point coefficient in~\eqref{eq:nondynCOT}, it is easy to see that the first term as well as the terms proportional to the complementary exponential integral and to $\gamma$ cancel out (notice that $z\to -z$ when $k_a\to -k_a$). To see the cancellation of the remaining terms, recall that the analytic continuation of $k_a$ is done via a rotation in the clockwise direction~\cite{Goodhew:2020hob}, so it accumulates a phase $-\pi$ in the argument of the logarithms,
\begin{equation}
    \ln (1/z) \quad\to\quad \ln (e^{-i\pi}/z) = \ln (1/z) -i\pi \, .
\end{equation}
The imaginary piece in~\eqref{eq:nondyn_pis4_Ein} precisely cancels these phases, and we conclude that the UV four-point wavefunction coefficient~\eqref{eq:nondynexactcalc} satisfies the COT. In the EFT, we have infinitely many contact diagrams $\psi_{4s}^{(n)}$. Each one of them individually satisfies the contact COT, which takes the form~\eqref{eq:nondynCOT}. Their resummation is identical to the result in the full theory with $\chi$ and so also obeys the COT in the form~\eqref{eq:nondynCOT}.


\section{Zero sound speed massive field}
\label{sec:zero cs}

The next case that we will study is that of a massive field $\chi$ with vanishing sound speed, i.e. a field whose equations of motion contains no spatial derivatives. The theory will also contain a conformally coupled scalar $\varphi$ and a cubic interaction between both. The Lagrangian is
\begin{equation} \label{eq:Lagr0speed}
    \mathcal{L} = - \frac{1}{2} a^4 (\partial_{\mu}\varphi)^2 - a^4 H^2 \varphi^2 + \frac{1}{2} a^2 (\chi')^2 - \frac{1}{2} M^2 a^4 \chi^2 + \lambda \, a^4 \varphi^2\chi \, .
\end{equation}
In contrast to the toy model we studied in the previous section, see~\eqref{eq:NonDynField}, the massive $\chi$ field is now dynamical and thus harder to integrate out. We focus on the less generic case of a vanishing speed of sound for $\chi$ because the absence of spatial derivatives makes the calculation manageable. This example constitutes a middle step between the very simple case studied in Section~\ref{sec:NonDynField} and the realistic but difficult case of a field with standard full kinetic term $(\nabla_{\mu}\chi)^2$. In this case, the EFT will keep track of just the $\varphi$, which, in contrast to the previous section, has a mass tuned to the conformally coupled valued rather than being massless. 


\subsection{Calculation in the full theory} \label{sec:cs0exact}

We start by computing the four-point correlator of $\varphi$ coming from tree-level exchange of $\chi$ in the full theory. We will do this using the field-theoretic wavefunction. The Bunch-Davies mode function and bulk-to-boundary propagator for the conformally coupled scalar are
\begin{equation}
    \varphi_k (\eta) \propto \eta e^{-ik\eta} \qquad\Rightarrow\qquad K^{\varphi}_k (\eta) = \frac{\eta}{\eta_0} e^{ik\eta} \, .
\end{equation}
The free (i.e. $\lambda=0$) equation of motion for the heavy field is
\begin{equation} \label{eq:HomEOM0speed}
    \Big[ \eta\partial_{\eta} (\eta\partial_{\eta}) + \mu^2 \Big] (\eta^{-3/2} \chi) = 0 \, , \qquad \text{where} \quad \mu^2 \ceq \frac{M^2}{H^2}-\frac{9}{4} \, .
\end{equation}
The solution is
\begin{equation}
    \chi (\eta) =\frac{H}{\sqrt{2\mu}} (-\eta)^\frac{3}{2} \left[ A \left( \frac{\eta}{\eta_*} \right)^{i\mu} + B \left( \frac{\eta}{\eta_*} \right)^{-i\mu}\right] \, ,
\end{equation}
where $-\eta_*$ is an arbitrary positive constant that makes the base of the exponents dimensionless and the constant prefactor is introduced to simplify later expressions. The normalization condition requires
\begin{equation} \label{eq:norm}
    \chi^*\chi' - (\chi')^*\chi = - i H^2\eta^2 \qquad\Rightarrow\qquad |A|^2-|B|^2 = 1\, .
\end{equation}
We will compute the wavefunction assuming generic boundary conditions for the field $\chi$, so $A$ and $B$ will only be constrained by~\eqref{eq:norm}. The reason why we keep these constants arbitrary deserves some explanation. First, to provide some context, we note that the limit of a vanishing sound speed is related to the so-called \textit{Carroll limit}, which describes a set of Wigner-Inonou contractions of kinematical groups \cite{Bacry:1968zf}. It is known that, in this limit, the quantization of fields is problematic \cite{deBoer:2023fnj}. Here the problem manifests itself in the ambiguity of choosing an initial state for $\chi$ that is the equivalent of the standard Bunch-Davies state. Since this toy model serves us as a concrete setup where calculations are manageable, rather than as a realistic construction, we prefer to simply keep $A$ and $B$ arbitrary and refrain from asking probing questions about the nature of the $\chi$ field and what initial conditions would be consistent or natural.

When we finally compare the EFT and exact theory it will, however, be interesting to demand that there are no heavy particles present in the initial state. To do this, we will take $A$ and $B$ to come from the zero speed of sound limit of the general massive-field solution in the Bunch-Davies vacuum,
\begin{align}
    \chi&=\lim_{c_s\rightarrow 0}\frac{\sqrt{\pi}H}{2}e^{-\frac{\pi\mu}{2}}\left(-\frac{\eta}{c_s}\right)^{3/2}H^{(1)}_{i\mu}(-c_s k\eta)\\&=\frac{H(-\eta)^{3/2}}{\sqrt{2\mu}}\left(-ie^{\frac{\pi\mu}{2}}\Gamma(-i\mu)\sqrt{\frac{\mu}{2\pi}}\left(-\frac{c_sk\eta}{2}\right)^{i\mu}-ie^{-\frac{\pi\mu}{2}}\Gamma(i\mu)\sqrt{\frac{\mu}{2\pi}}\left(-\frac{c_sk\eta}{2}\right)^{-i\mu}\right).
\end{align}
This defines for us a specific value of $A$ and $B$ as well as an $\eta_*$,
\begin{align}\label{eq:B-D}
    A&=-e^{\frac{\pi\mu}{2}}\Gamma(-i\mu)\sqrt{\frac{\mu}{2\pi}},& B&=-ie^{-\frac{\pi\mu}{2}}\Gamma(i\mu)\sqrt{\frac{\mu}{2\pi}},&\eta_*&=-\frac{2}{c_sk}.
\end{align}
However, for the majority of this section, this specific choice will not be assumed. The bulk-to-boundary and bulk-to-bulk propagators are
\begin{align} 
    K^{\chi} (\eta) & = \frac{\chi^*(\eta)}{\chi^*(\eta_0)} \, , \\
    G^{\chi} (\eta,\eta') & = \chi (\eta) \chi^* (\eta') \theta(\eta-\eta') + \chi^* (\eta) \chi (\eta') \theta(\eta'-\eta) - \chi^* (\eta) \chi^* (\eta') \frac{\chi (\eta_0)}{\chi^* (\eta_0)} \, , \label{eq:bulktobulk}
\end{align}
and satisfy the boundary conditions $K^\chi(\eta_0)=1$ and $G^{\chi} (\eta_0,\eta')=0$ at some late time $\eta_0$. 


\paragraph{2- and 3-point functions}

Using~\eqref{eq:psi2grformula} we find that the two-point wavefunction coefficients for $\varphi$ and $\chi$ are respectively
\begin{align}
    \psi_2^{\varphi} & = - \frac{i}{H^2\eta_0^3} (1+ik\eta_0) \, , \\
    \psi_2^{\chi} & = - \frac{i}{H^2\eta_0^3} \left[ \frac{3}{2} - i\mu \frac{A^*(\eta_0/\eta_*)^{-i\mu}-B^*(\eta_0/\eta_*)^{i\mu}}{A^*(\eta_0/\eta_*)^{-i\mu}+B^*(\eta_0/\eta_*)^{i\mu}} \right] \, .
\end{align}
The three-point wavefunction coefficient coming from the cubic interaction in~\eqref{eq:Lagr0speed} is
\begin{equation} \label{eq:cs0psi3}
    \psi_3 = i\lambda \int_{-\infty}^0 \dd \eta \, a(\eta)^4 \, K^{\varphi}_{k_1} (\eta) \, K^{\varphi}_{k_2} (\eta) \, K^{\chi} (\eta) = \frac{i\lambda}{H^4(-\eta_0)^{7/2}} \frac{A^* \, I^{(1)}_- (k_{12}) + B^* \, I^{(1)}_+ (k_{12})}{A^* ( \eta_0/\eta_* )^{-i\mu} + B^* ( \eta_0/\eta_* )^{i\mu}} \, ,
\end{equation}
where $k_{ab}\ceq k_a+k_b$ and we have defined
\begin{equation} \label{eq:int1}
    I^{(1)}_{\pm} (k) \ceq (-\eta_*)^{\mp i\mu} \int_{-\infty}^0 \dd \eta \, (-\eta)^{-\frac{1}{2}\pm i\mu} \, e^{ik\eta} = (-\eta_*)^{\mp i\mu} \frac{ e^{i\frac{3\pi}{4} \pm\frac{\pi\mu}{2}} }{ k^{\frac{1}{2}\pm i\mu} } \, \Gamma \left( \frac{1}{2} \pm i\mu \right) \, .
\end{equation}


\paragraph{4-point function}

The $s$-channel exchange wavefunction coefficient is:
\begin{equation} \label{eq:cs04pt}
    \psi_{4s} = - \lambda^2 \int_{-\infty}^0 \dd \eta \, \dd \eta' \, a(\eta)^4 \, a(\eta')^4 \, K^{\varphi}_{k_1} (\eta) \, K^{\varphi}_{k_2} (\eta) \, G^{\chi} (\eta,\eta') \, K^{\varphi}_{k_3} (\eta') \, K^{\varphi}_{k_4} (\eta') \, .
\end{equation}
We separate the contributions from the three terms in~\eqref{eq:bulktobulk} as
\begin{equation}
    \psi_{4s} = \psi_{4s}^{({\rm I})} + \psi_{4s}^{({\rm II})} + \psi_{4s}^{({\rm III})} \, .
\end{equation}
The first time-ordered term gives
\begin{equation}
\begin{aligned}
    \psi_{4s}^{({\rm I})}(k_{12},k_{34}) = \frac{-\lambda^2}{2\mu H^6\eta_0^4} \Bigg[ & |A|^2 I^{(2)}_{+-} (k_{12},k_{34}) + |B|^2 I^{(2)}_{-+} (k_{12},k_{34}) \\
    & + A^*B \, I^{(2)}_{--} (k_{12},k_{34}) + AB^* I^{(2)}_{++} (k_{12},k_{34}) \Bigg] \, ,
\end{aligned}
\end{equation}
with 
\begin{equation} \label{eq:int2}
\begin{aligned}
    I^{(2)}_{\pm\mp} (k,k') & \ceq \int_{-\infty}^0 \dd \eta \, (-\eta)^{-\frac{1}{2}\pm i\mu} \, e^{ik\eta} \int_{-\infty}^{\eta} \dd \eta' (-\eta')^{-\frac{1}{2}\mp i\mu} \, e^{ik'\eta'} \\
    & = -\frac{i}{\left( \frac{1}{2} \pm i\mu \right) (k+k') } \, {}_2F_1 \left( 1, 1 ; \frac{3}{2} \pm i\mu ; \frac{k}{k+k'} \right) \, , \\[5pt]
    I^{(2)}_{\pm\pm} (k,k') & \ceq (-\eta_*)^{\mp i2\mu} \int_{-\infty}^0 \dd \eta \, (-\eta)^{-\frac{1}{2}\pm i\mu} \, e^{ik\eta} \int_{-\infty}^{\eta} \dd \eta' (-\eta')^{-\frac{1}{2}\pm i\mu} \, e^{ik'\eta'} \\
    & =-(-\eta_*)^{\mp i2\mu}\frac{i \, e^{\pm\pi\mu} \, \Gamma \left( 1 \pm 2i\mu \right)}{\left( \frac{1}{2} \pm i\mu \right) (k+k')^{1 \pm 2i\mu} } \, {}_2F_1 \left( 1, 1 \pm 2i\mu ; \frac{3}{2} \pm i\mu ; \frac{k}{k+k'} \right) \, , 
\end{aligned}
\end{equation}
and the second gives the same result with $k_{12}$ and $k_{34}$ swapped,
\begin{equation}
    \psi_{4s}^{({\rm II})}(k_{12},k_{34}) = \psi_{4s}^{({\rm I})} (k_{34},k_{12}) \, .
\end{equation}
Using the properties of the hypergeometric functions~\cite{NIST:DLMF} we can see that
\begin{align}
    I^{(2)}_{\pm\pm}(k,k')+I^{(2)}_{\pm\pm}(k',k)&=I^{(1)}_{\pm}(k)I^{(1)}_{\pm}(k'),\\
    I^{(2)}_{\pm\mp}(k,k')+I^{(2)}_{\mp\pm}(k',k)&=I^{(1)}_{\pm}(k)I^{(1)}_{\mp}(k').
\end{align}
Therefore,
\begin{align}\label{eq:PsiExact}
    \psi_{4s}^{({\rm I})}+\psi_{4s}^{({\rm II})} &= \frac{-\lambda^2}{2\mu H^6\eta_0^4} \Bigg[   I^{(2)}_{+-} (k_{12},k_{34})+I^{(2)}_{+-} (k_{34},k_{12})+A^*B  \, I^{(1)}_{-}(k_{12})I^{(1)}_{-}(k_{34})  \\
    & + AB^*\, I^{(1)}_{+}(k_{12})I^{(1)}_{+}(k_{34}) +|B|^2\left( I^{(1)}_-(k_{12})I^{(1)}_+(k_{34})+I^{(1)}_-(k_{34})I^{(1)}_+(k_{12}) \right)\Bigg] \, , \nonumber
\end{align}
where we have employed the relationship~\eqref{eq:norm}. 
The non-time-ordered term gives
\begin{align} \label{eq:PsiExactIII}
    \psi_{4s}^{({\rm III})} = - \frac{\psi_3 (k_{12}) \, \psi_3 (k_{34}) }{2\, \text{Re} \, \psi_2^{\chi}} \, .
\end{align}


\paragraph{Correlator}
The correlation function of four conformally coupled scalars $\varphi$ exchanging a massive $\chi$ in the $s$-channel is given by 
\begin{equation}
\begin{aligned}
    & \langle \varphi_{\mathbf{k}_1}\varphi_{\mathbf{k}_2}\varphi_{\mathbf{k}_3}\varphi_{\mathbf{k}_4} \rangle_s = (2\pi)^3 \delta(\sum_{a=1}^4\bfk_a) B_{4s} \,, \qquad\text{with} \\
    B_{4s}&=\left( \prod_{a=1}^4 2\, {\rm Re} \, \psi_2^{\varphi} (k_a) \right)^{-1} \left[ 2\, {\rm Re} \, \psi_{4s} (k_{12},k_{34}) + \frac{2\, {\rm Re} \, \psi_3 (k_{12}) \, 2\, {\rm Re} \, \psi_3 (k_{34})}{2\, {\rm Re} \, \psi_2^{\chi}} \right] \\
    & = \left( \prod_{a=1}^4 2\, {\rm Re} \, \psi_2^{\varphi} (k_a) \right)^{-1} \left[ 2\, {\rm Re} \, \big( \psi_{4s}^{({\rm I})} + \psi_{4s}^{({\rm II})} \big) + \frac{{\rm Re} \, \big( \psi_3 (k_{12}) \, \psi_3^* (k_{34}) \big)}{{\rm Re} \, \psi_2^{\chi}} \right] \, .
\end{aligned}
\end{equation}
Substituting the wavefunction coefficients obtained above we get
\begin{equation} \label{eq:B4correl}
\begin{aligned}
    B_{4s} & = \frac{H^2\lambda^2\eta_0^4}{32 \, \mu k_1k_2k_3k_4} \Bigg\{ - 2 \, {\rm Re} \Big( I^{(2)}_{+-} (k_{12},k_{34})+I^{(2)}_{+-} (k_{34},k_{12}) \Big) \\
    & + |A|^2 \, 2 \, {\rm Re} \Big( I^{(1)}_- (k_{12}) \big( I^{(1)}_- (k_{34}) \big)^* \Big) + |B|^2 \, 2 \, {\rm Re} \Big( I^{(1)}_+ (k_{12}) \big( I^{(1)}_+ (k_{34}) \big)^* \Big) \\
    &  - A^*B \Big[ I^{(1)}_- (k_{12})  - \big( I^{(1)}_+ (k_{12}) \big)^* \Big] \Big[ I^{(1)}_- (k_{34})  - \big( I^{(1)}_+ (k_{34}) \big)^* \Big] \\
    & - AB^* \Big[ I^{(1)}_+ (k_{12}) -\big( I^{(1)}_- (k_{12}) \big)^* \Big] \Big[ I^{(1)}_+ (k_{34})  - \big( I^{(1)}_- (k_{34}) \big)^* \Big]  \Bigg\} \, .
\end{aligned}
\end{equation}
One can verify that the same result is found when the correlator is calculated in the in-in formalism.
In that case, there is a factorized contribution $B_{4s}^\text{fact}$ arising from diagrams with $G_{\pm \mp}$ propagators of the heavy $\chi$ field, and a non-factorized contribution $B_{4s}^\text{non-fact}$ from diagrams with $G_{\pm \pm}$ propagators for $\chi$,
\begin{align}
    B_{4s}&=B_{4s}^\text{fact}+B_{4s}^\text{non-fact}\,.
\end{align}
These contributions are
\begin{align}
   B_{4s}^{\text{fact}} = & \,
   \frac{H^2\lambda^2 \eta_0^4}{32\mu k_1 k_2 k_3 k_4}  \Big[ \abs{A}^2 2 \,\Re\left( I^{(1)}_-(k_{12}) I^{(1)}_{-}(k_{34})^* \right) + \abs{B}^2 2 \,\Re\left( I^{(1)}_+(k_{12}) I^{(1)}_{+}(k_{34})^* \right) 
    \nonumber \\ &
      + A^* B \left( I^{(1)}_-(k_{12}) I^{(1)}_+(k_{34})^* + I^{(1)}_+(k_{12})^* I^{(1)}_-(k_{34}) \right)
    \nonumber \\ & 
      + A B^* \left( I^{(1)}_+(k_{12}) I^{(1)}_-(k_{34})^* + I^{(1)}_-(k_{12})^* I^{(1)}_+(k_{34}) \right)
   \Big]  \label{eq:in-in trisp fact} \, ,\\
B_{4s}^\text{non-fact} = & \, 
\frac{H^2\lambda^2 \eta_0^4}{32\mu k_1 k_2 k_3 k_4} \Big[ -2 \Re\left( I^{(2)}_{+-}(k_{12},k_{34}) + I^{(2)}_{+-}(k_{34},k_{12}) \right) \nonumber \\ & 
      -  A^* B \left( I_-^{(1)}(k_{12}) I_-^{(1)}(k_{34}) + I_+^{(1)}(k_{12})^* I_+^{(1)}(k_{34})^* \right) \nonumber \\ & 
      - A B^* \left( I_+^{(1)}(k_{12}) I_+^{(1)}(k_{34}) + I_-^{(1)}(k_{12})^* I_-^{(1)}(k_{34})^* \right) 
   \Big] \, .
   \label{eq:in-in trisp non-fact}
\end{align}
We now proceed to discussing the EFT in which $\chi$ is integrated out.


\subsection{Calculation in the EFT and resummation}\label{sec:EFT2}

Before we get into the specifics of this model we will review the general properties of EFTs in the presence of a boundary, discussed in~\cite{salcedo2023analytic}, as there are some subtleties that were not important in Section~\ref{sec2:EFT} but that will arise here. Suppose we have some theory with a massive and massless field that are coupled through the following Lagrangian
\begin{align}
    S=\int_\bfx\int \dd\eta \left( -\frac{1}{2}a^4\partial_\mu\phi\partial^\mu\phi +\frac{1}{2}a^2\partial_\eta\chi\partial_\eta\chi-\frac{1}{2}c_s^2 a^2\partial_i\chi\partial_i\chi-\frac{1}{2}Ma^4\chi^2 +\lambda \mathcal{L}_{\text{int}}[\chi,\phi] \right) \, .
\end{align}
We have temporarily restored the spatial derivatives in this theory and introduced an arbitrary sound speed for the purpose of this general discussion. When we return to discussing the zero speed of sound limit, all the arguments made here will continue to apply. Suppose we wish to integrate out the massive field, $\chi$, then we must solve its equations of motion
\begin{align}
    \left(\partial_\eta a^2\partial_\eta-c_s^2a^2\partial_i\partial_i+a^4M^2\right)\chi\eqc\mathcal{O}\,\chi=\lambda \frac{\partial\mathcal{L}_{\text{int}}}{\partial \chi}.
\end{align}
This has formal solution
\begin{align}
    \chi=\lambda\left(\left.\mathcal{O}^{-1}\frac{\partial\mathcal{L}_{\text{int}}}{\partial \chi}\right\rvert_{\chi=0}+\chi_{\text{CF}}\right)\,,
\end{align}
where $\chi_{\text{CF}}$ is a solution to the homogeneous equations of motion that is required to fix the boundary conditions that we impose on $\chi$. The fact that $\chi_{\text{CF}}$ appears with a $\lambda$ factor in the above expression is due to the fact that we are working in the EFT where the background value of $\chi$ vanishes.
Putting this solution back into our action gives the effective action which we can integrate by parts to make the operator that defines our equations of motion manifest
\begin{align}\nonumber\label{eq:SEFT}
    S&=\int_\bfx \Bigg[\frac{\lambda^2}{2}\left(\mathcal{O}^{-1}\frac{\partial\mathcal{L}_{\text{int}}}{\partial \chi}+\chi_{\text{CF}}\right)a^2\partial_\eta \left(\mathcal{O}^{-1}\frac{\partial\mathcal{L}_{\text{int}}}{\partial \chi}+\chi_{\text{CF}}\right) { \Bigg]^{\eta_0}_{-\infty}} +\int_\bfx \int \dd\eta  \Bigg[ \frac{1}{2}\nabla_\mu\phi\nabla^\mu\phi +\lambda \mathcal{L}_{\text{int}}\left[0,\phi\right]\\
    &-\frac{\lambda^2}{2}\left(\mathcal{O}^{-1}\frac{\partial\mathcal{L}_{\text{int}}}{\partial \chi}+\chi_{\text{CF}}\right)\mathcal{O}\left(\mathcal{O}^{-1}\frac{\partial\mathcal{L}_{\text{int}}}{\partial \chi}+\chi_{\text{CF}}\right) +\lambda^2\left(\mathcal{O}^{-1}\frac{\partial\mathcal{L}_{\text{int}}}{\partial \chi}+\chi_{\text{CF}}\right)\frac{\partial\mathcal{L}_{\text{int}}}{\partial\chi} \Bigg] \, .  
\end{align}
In this expression we have dropped the notation that $\mathcal{L}_{\text{int}}$ is evaluated at $\chi=0$ and expanded the final expression for $\mathcal{L}_{\text{int}}$ assuming $\lambda$ is small. To simplify the second line we first note that $\mathcal{O}\chi_{\text{CF}}=0$. Therefore, we can see that, when we cancel the $\mathcal{O}$ and $\mathcal{O}^{-1}$ the first term is equal to the second one. As in~\cite{salcedo2023analytic} this can be further simplified by integration by parts,
\begin{align}\label{eq:IBP}
    {\frac{\lambda^2}{2}} \int_\bfx \int \dd\eta \,\chi_{\text{CF}}\mathcal{O}\mathcal{O}^{-1}\frac{\partial\mathcal{L}_{\text{int}}}{\partial\chi}= {\frac{\lambda^2}{2}} \int_\bfx \left[\chi_{\text{CF}}a^2\partial_\eta \mathcal{O}^{-1}\frac{\partial\mathcal{L}_{\text{int}}}{\partial \chi}-\mathcal{O}^{-1}\frac{\partial\mathcal{L}_{\text{int}}}{\partial \chi}a^2\partial_\eta  \chi_{\text{CF}}\right]^{\eta_0}_{-\infty}.
\end{align}
Here there is no bulk integral term on the right hand side as this procedure has moved the operator $\mathcal{O}$ onto $\mathcal{\chi}_{\text{CF}}$, which therefore vanishes.

We now return to the theory at hand. The first step is to determine the EFT operators that enter the theory. To do this we need the equations of motion which we manipulate into a simple form,
\begin{align}
    \left(\frac{1}{\mu^2}\eta \partial_\eta(\eta \partial_\eta)+1\right)\left((-\eta)^{-3/2}\chi\right)=\frac{\lambda}{H^2\mu^2}(-\eta)^{-3/2}\varphi^2.
\end{align}
This differential operator can be inverted and expanded to give an expression for $\chi$ in terms of an infinite sum of local-in-time operators
\begin{align}\label{eq:chisol}
    {\chi}=\frac{\lambda(-\eta)^{3/2}}{H^2\mu^2}\sum_{n=0}^\infty \left(-\frac{1}{\mu^2}\right)^{n}(\eta\partial_\eta)^{2n}\left((-\eta)^{-3/2}\varphi^2\right)+\lambda{\chi}_{\text{CF}} \, .
\end{align}
The Lagrangian recovered upon integrating out the $\chi$ field can thus be read off from~\eqref{eq:SEFT} and~\eqref{eq:IBP}
\begin{align}\nonumber
   S&=  S_{\text{BD}}+S_{\text{EFT}}\\ 
    &={\int_\bfx \,}\left[\frac{1}{2}a^2 \chi \partial_\eta \chi+\lambda a^2\chi_{\text{CF}}\partial_\eta \chi -\lambda a^2\chi \partial_\eta \chi_{\text{CF}}\right]_{-\infty}^{\eta_0}+\int_{\bfx}\int \dd\eta\Bigg[ \frac{1}{2}a^2\varphi'^2 -\frac{1}{2}a^2 \partial_i\varphi\partial^i\varphi \label{334} \\ 
    & \quad - H^2 a^4 \varphi^2 +\frac{\lambda^2}{2\mu^2H^6(-\eta)^{5/2}}\varphi^2\sum_{n=0}^\infty \left(-\frac{1}{\mu^2}\eta\partial_\eta(\eta\partial_\eta)\right)^n\left(\varphi^2 (-\eta)^{-3/2}\right) \Bigg]. \nonumber
\end{align}
Here we have expressed the boundary contribution in terms of the full solution to keep the expression concise. We will ultimately be evaluating this using the infinite sum in~\eqref{eq:chisol}.

To compute the wavefunction from this expression we need to know the action of the differential operator $(\eta\partial_\eta)^{2n}$ on the conformally coupled mode functions. In general one has the identity
\begin{align}
    (\eta\partial_\eta)^{2n}(-\eta)^m e^{ik\eta}=(-\eta)^me^{ik\eta}\sum_{b=0}^{2n}\sum_{a=0}^b(-1)^{a+b}\binom{b}{a}\frac{(a+m)^{2n}}{b!}(ik\eta)^b\,.
\end{align}
In our case the conformally coupled mode functions are proportional to $\eta e^{ik\eta}$ and so $m=\frac{1}{2}$.

By computing the time integral in the Feynman rules for the contact wavefunction coefficient generated by a generic interaction in~\eqref{334} we find
\begin{equation} \label{eq:cs0psi4n}
\begin{aligned}
    \psi_{4s}^{(n)}(k_{12},k_{34})&=\frac{\lambda^2}{2H^6\mu^2 \eta_0^4 k_T}\sum_{b=0}^{2n}\sum_{a=0}^b \left(-\frac{1}{\mu^2}\right)^n(-1)^{a}\binom{b}{a}\left(\frac{k_{12}}{k_T}\right)^b\left(a+\frac{1}{2}\right)^{2n}.
\end{aligned}
\end{equation}
This double sum can be replaced by a single sum as
\begin{align}\label{eq:flipsum}
    \psi_{4s}^{(n)}(k_{12},k_{34})=\frac{\lambda^2}{2H^6\mu^2\eta_0^4}\sum_{a=0}^\infty \left(-\frac{1}{\mu^2}\right)^n \frac{1}{k_{34}}\left(-\frac{k_{12}}{k_{34}}\right)^a\left(a+\frac{1}{2}\right)^{2n}\eqc \frac{\lambda^2}{H^6\mu^2\eta_0^4}\sum_{a=0}^\infty A^{(n)}_a \, .
\end{align}
The full EFT expression then comes from summing over all values $n$. However, this sum does not converge. We can produce a resummed expression for this EFT if we instead exchange the order of summation:
\begin{equation}
\begin{aligned}
    \psi_{4s}^{\text{EFT}} &=\sum_{n=0}^\infty \left[ \psi_{4s}^{(n)}(k_{12},k_{34})+ \psi_{4s}^{(n)}(k_{34},k_{12})\right] =\frac{\lambda^2}{2H^6\mu^2\eta_0^4}\sum_{n=0}^\infty \sum_{a=0}^\infty  A^{(n)}_a +(k_{12}\leftrightarrow k_{34}) \\
    & = \frac{\lambda^2}{2H^6\mu^2\eta_0^4}\sum_{a=0}^\infty \sum_{n=0}^\infty  A^{(n)}_a+(k_{12}\leftrightarrow k_{34}),
\end{aligned}
\end{equation}
where we have made the two contributions to this expression in the $s$ channel explicit. To make sense of this double sum we propose a resummation procedure as follows. First we look at the sum over $n$ at fixed $a$. We apply the ratio test to the sum over $n$, finding the condition
\begin{align}
    \lim_{n\rightarrow \infty} \frac{|A_{a}^{(n+1)}|}{|A_{a}^{(n)}|}=\frac{(a+1/2)^2}{\mu^2} <1\,.
\end{align}
For any finite $a$, not necessarily integer, we can find a finite range of $\mu$ for which this sum converges to
\begin{align}
    \sum_{n=0}^\infty A_a^{(n)} =\frac{4\mu^2}{k_{34}(4\mu^2+(1+2a)^2)}\left(-\frac{k_{12}}{k_{34}}\right)^a.
\end{align}
Now we can analytically continue this expression in $a$ to arbitrary values. Then we can perform the sum over $a$ which now converges to 
\begin{align} \label{eq:psi4 EFT}
    \psi_{4s}^{\text{EFT}}=-\frac{ \lambda^2}{2\mu H^6\eta_0^4}\left(I^{(2)}_{+-}(k_{12},k_{34})+I^{(2)}_{+-}(k_{34},k_{12})-\frac{1}{2}I^{(1)}_-(k_{12})I^{(1)}_+(k_{34})-\frac{1}{2}I^{(1)}_-(k_{34})I^{(1)}_+(k_{12})\right) \, ,
\end{align}
where we have written the resulting hypergeometric functions in terms of the integrals defined in~\eqref{eq:int2} for ease of comparison. This makes it clear that for any boundary conditions the first two terms are identical to those in the exact result in~\eqref{eq:PsiExact}. Conversely, to match the remaining terms we would need to consider the potential additional boundary term ${\chi}_{\text{CF}}$, as we show in Appendix~\ref{app:psiBC}. This requires information that exists outside of the EFT as the form of ${\chi}_{\text{CF}}$ depends on the solution of the full theory. 

One can ask how close the EFT resummed expression in~\eqref{eq:psi4 EFT} approximates the true result of the full theory, without invoking the knowledge of any boundary terms. To make this comparison we look at the infinite mass limit of the difference between the EFT and exact calculations,
\begin{align}
    \lim_{\mu\rightarrow\infty}\psi_{4s}-\psi_{4s}^{\text{EFT}}=&\,-\frac{ \lambda^2}{2\mu H^6\eta_0^4\sqrt{k_{12}k_{34}}}\Bigg[\frac{iB^*(\eta_*^2k_{34}k_{12})^{-i\mu}}{A^*(\eta_0/\eta_*)^{-i\mu}+B^*(\eta_0/\eta_*)^{i\mu}}\left(\frac{\eta_0}{\eta_*}\right)^{i\mu}\Bigg]+\mathcal{O}(e^{-\pi\mu})\,,
\end{align}
where we have used that 
\begin{align}
    \lim_{\mu\rightarrow \infty} I_+^{(1)}(k)&=(-\eta_*)^{-i\mu} \, k^{-\frac{1}{2}- i\mu} e^{-i\frac{3\pi}{4}},\\
    \lim_{\mu\rightarrow \infty} I_-^{(1)}(k)&=e^{-\pi\mu}(-\eta_*)^{i\mu} \, k^{-\frac{1}{2}+ i\mu} e^{i\frac{\pi}{4}}.
\end{align}
to conclude that anything involving $I_-^{(1)}$ is exponentially suppressed and hence $\mathcal{O}(e^{-\pi\mu})$. There is still a non-exponentially suppressed contribution to this wavefunction coefficient for generic values of $A$ and $B$.

However, taking the values of $A$ and $B$ to coincide with the general massive-field solution in the Bunch-Davies vacuum, given in~\eqref{eq:B-D}, we find that
\begin{align}
    \lim_{\mu\rightarrow\infty} \frac{B^*}{A^*}=\lim_{\mu\rightarrow\infty}\frac{\pi(1+\coth(\pi\mu))}{\mu\Gamma(i\mu)^2}=\frac{i}{2}\left(\frac{e}{m}\right)^{2i\mu}e^{\pi\mu}(1+\coth(\pi\mu))=\frac{i}{2}\left(\frac{e}{m}\right)^{2i\mu}e^{-\pi\mu}.
\end{align}
These boundary conditions thus ensure that
\begin{align}
    \lim_{\mu\rightarrow\infty}\psi_{4s}-\psi_{4s}^{\text{EFT}}=\mathcal{O}(e^{-\pi\mu}).
\end{align}
The same is true for the correlator as the three point wavefunction coefficient also vanishes exponentially fast in this limit
\begin{align}
    \lim_{\mu\rightarrow\infty}\psi_3=\lim_{\mu\rightarrow\infty}\frac{i\lambda}{H^4(-\eta_0)^{7/2}}\frac{B^*(-\eta_*k)^{-i\mu}e^{-\frac{3i\pi}{4}}}{A^*(\eta_0/\eta_*)^{-i\mu}+B^*(\eta_0/\eta_*)^{i\mu}}=\mathcal{O}(e^{-\pi\mu}).
\end{align}
Therefore, the EFT recovers the true solution up to exponentially small corrections whether we are looking at the wavefunction coefficient or the correlator. It may initially seem concerning that we were required to specify a particular initial condition for $\chi$ to match the exact result as the EFT shouldn't have access to this information, which pertains only to the full theory. However, if the initial state of the system contains some on-shell $\chi$ particles, rather than being in the Fock vacuum, then the EFT prescription that we have implemented of setting the heavy field to zero will differ from the true description in a way that is not suppressed in the infinite mass limit and so we expect that the mistake the EFT makes as compared to the result of the full theory is not necessarily exponentially suppressed in the large mass limit. 


\subsection{Relationship to the Cosmological Optical Theorem} \label{sec:zerocsCOT}

We can look at the consequences of unitarity in this theory just like we did for the previous one in Section~\ref{sec:nondynCOT}. This case is more subtle, however, because we work with generic initial conditions for the massive field $\chi$. For a three-point wavefunction coefficient in a scalar theory, the generic constraint from unitarity is~\cite{Cespedes:2020xqq}\footnote{One could alternatively use the cutting rules for Bogoliubov initial states derived in~\cite{Ghosh:2024aqd,Ghosh:2025pxn}.}
\begin{equation}
    \psi_3(k_a)+[\psi_3(\bar{k}_a)]^* = \beta_3 (k_a) \, ,
\end{equation}
where $a=1,2,3$, the bar in $\bar{k}$ denotes the analytic continuation of the momentum that achieves $f_{\bar{k}}=f^*_{k}$ in the corresponding mode function, and $\beta_3 (k_a)$ is a function of the momenta that vanishes if the three fields are taken to be in the Bunch-Davies vacuum. In our case the three-point function~\eqref{eq:cs0psi3} only depends on the momenta of the c.c. fields via $k_{12}$, so the above expression simplifies to
\begin{equation} \label{eq:cs0psi3disc}
    \psi_3(k_{12})+[\psi_3(-k_{12}^*)]^* = \beta (k_{12}) = \frac{i\lambda}{H^4(-\eta_0)^{7/2}} \frac{( \eta_0/\eta_* )^{i\mu} \, I^{(1)}_- (k_{12}) - ( \eta_0/\eta_* )^{-i\mu} \, I^{(1)}_+ (k_{12})}{\left| A ( \eta_0/\eta_* )^{i\mu} + B ( \eta_0/\eta_* )^{-i\mu} \right|^2} \, ,
\end{equation}
where in the last equality we have plugged in the explicit result~\eqref{eq:cs0psi3}. 

For the $s$-channel four-point function, unitarity implies~\cite{Cespedes:2020xqq}
\begin{equation}
    \psi_{4s}(k_a,s)+[\psi_{4s}(\bar{k}_a,s)]^* = \frac{\left[ \psi_3 (k_1,k_2,s) + [\psi_3 (\bar{k}_1,\bar{k}_2,s)]^* \right] \left[ \psi_3 (k_3,k_4,s) + [\psi_3 (\bar{k}_3,\bar{k}_4,s)]^* \right] }{2\,{\rm Re}\,\psi_2 (s)} + \beta_{4s} (k_a,s) \, ,
\end{equation}
where $a=1,\ldots,4$. In our theory, and focusing on the coefficient~\eqref{eq:cs04pt}, the four external fields are conformally coupled scalars with BD initial condition, so we have $\beta_{4s} (k_a)=0$. In addition, the three-point discontinuities in the right hand side are $\psi_3 (k_1,k_2,s) + [\psi_3 (\bar{k}_1,\bar{k}_2,s)]^* = \psi_3(k_{12})+[\psi_3(-k_{12}^*)]^*$, so we can use~\eqref{eq:cs0psi3disc} getting
\begin{equation} \label{eq:cs0COT}
    \psi_{4s}(k_{12},k_{34})+[\psi_{4s}(-k_{12}^*,-k_{34}^*)]^* = \frac{\beta_3 (k_{12}) \beta_3 (k_{34}) }{2\,{\rm Re}\,\psi_2^{\chi}} \, .
\end{equation}
One can easily check that this constraint is satisfied by the UV four-point wavefunction coefficient computed in~\eqref{eq:PsiExact} and~\eqref{eq:PsiExactIII}. However, the four-point coefficients $\psi_{4s}^{(n)}$ of~\eqref{eq:cs0psi4n} are computed in the EFT where the massive field has been integrated out, and come from contact interactions. As a consequence, they do not satisfy~\eqref{eq:cs0COT}, but its contact version, which has a vanishing right-hand side. This is also the case after resummation,~\eqref{eq:psi4 EFT}, i.e.
\begin{equation}
    \psi_{4s}^{\rm EFT} (k_a) + [\psi_{4s}^{\rm EFT} (-k_a^*)]^* = 0 \, .
\end{equation}
We conclude that the EFT misses the information about the discontinuity in the full theory, that is, about the right-hand side of~\eqref{eq:cs0COT}. This information is encoded in the complementary function terms studied in Appendix~\ref{app:psiBC}, which require knowledge of the UV theory.


\section{Mellin-space resummation at zero sound speed}
\label{sec:Mellin}
By studying the EFT derivative expansion in the zero speed of sound theory via Mellin space, an alternative method of resummation will present itself, with a finite radius of convergence, analogous to the case of flat-space scattering amplitudes described in Section~\ref{IntroSect}.
The Mellin transform $f(s)$ of a function $f(\eta)$ defined in conformal time is given by
\begin{equation}
   (\mcM f)(s) = f(s) = \int_{-\infty}^0 \frac{\dd \eta}{-\eta} (-\eta)^{2s} f(\eta) \, ,
\end{equation}
where $s$ is a complex variable, corresponding to a projection onto a basis of complex power-law functions in conformal time.
This diagonalizes scalings in conformal time, and because $-\eta = e^{-H t}/H$, where $t$ is cosmic time,
\begin{equation}
   f(s) = \int_{-\infty}^\infty H \dd t  \,e^{-2 H s t} f(t) \, .
\end{equation}
Therefore, the Mellin transform is a two-sided Laplace or Fourier transform in cosmic time.
Integrals in Mellin space often take the form of so-called Mellin-Barnes integrals, which are of the general form\footnote{This family of integrals is not sharply defined; an important characteristic is that there are gamma functions whose poles are separated by the contour of integration.}
\begin{equation}
   I(z) = \int_{c-i\infty}^{c+i\infty} [\dd s] \, \frac{\prod_i \Gamma\left( a_i s + \alpha_i  \right)}{\prod_i \Gamma\left( b_i s + \beta_i  \right)} z^{-2s} \, ,
\end{equation}
where $[\dd s] \ceq \dd s/2\pi i$ and the contour of integration separates the poles of the gamma functions with $a_i > 0$ from those with $a_i < 0$.
For a general introduction to this integral transform and its properties, see \cite{Paris_Kaminski_2001}.
Mellin-space methods have been applied to a broad range of problems in cosmology \cite{Sleight:2019mgd,Sleight:2019hfp,Qin:2022lva,Qin:2024gtr,Green:2024cmx,Pimentel:2025rds}, holography \cite{Fitzpatrick:2011ia,Alaverdian:2024llo}, and particle physics \cite{Dubovyk:2022obc}.
For some relevant properties of the Mellin transform and a formulation of the Feynman rules for cosmological correlators in Mellin space, see Appendix~\ref{app:Mellin}.
An equivalent diagrammatic formalism appears already in the Mellin space literature \cite{Sleight:2021plv, Sleight:2021iix}.


\subsection{Calculation in the EFT and resummation}

In this section, we will study correlators rather than wavefunction coefficients: this will also illustrate the extent to which a unitary effective field theory can successfully reproduce observables of the UV theory.
We will study the $s$-channel contribution to the trispectrum 
of the zero sound speed model of Section~\ref{sec:zero cs}.
In that calculation, the effective action \eqref{334} for the wavefunction was written in terms of time derivatives acting on $(-\eta)^{3/2}\varphi^2$ in order to make the calculation tractable.
In Mellin space, this is not required, and we write exactly the same interactions in a slightly different form
\begin{align}
   H_\text{int} &\ceq \sum_n \varphi^2 \mcO_n \varphi^2 \qquad \text{where} \qquad
   \mcO_n \ceq \frac{{-}\lambda^2}{4 M^2} \left[ -\frac{H^2}{M^2} \left( \eta\partial_\eta(\eta\partial_\eta) - 3 \eta\partial_\eta \right) \right]^n \, .
\end{align}
These interactions can be understood analogously to the flat-space case \eqref{eq:A4EFT}: the Feynman propagator of the heavy field $\chi$ is
\begin{equation}
   G^\chi_F = \frac{1}{H^2\left( \eta\partial_\eta (\eta\partial_\eta) - 3 \eta\partial_\eta \right) + M^2} \, ,
\end{equation}
and we aim to describe the trispectrum of $\varphi$ produced by the $\lambda \chi\varphi^2/2$ using a unitary effective description in terms of the conformally-coupled field $\varphi$ alone.
To that end, we Taylor expand the propagator of $\chi$ assuming large mass:
\begin{equation}
   -\frac{\lambda^2}{4}\varphi^2  \frac{1}{H^2\left( \eta\partial_\eta (\eta\partial_\eta) - 3 \eta\partial_\eta \right) + M^2} \varphi^2 \approx H_\text{int} \, .
\end{equation}
The EFT expression for the contact trispectrum in the $s$-channel then takes the form
\begin{equation}\label{timeinte}
   B_{4s}^\EFT = \sum_{n=0}^\infty 2 \, \Re  \int \frac{-i\dd \eta}{(-H\eta)^4} G_+(k_1, \eta) G_+(k_2, \eta) \mcO_n \left[ G_+(k_3, \eta) G_+(k_4, \eta) \right] { + (k_1,\,k_2\leftrightarrow k_3, \,k_4)} \, ,
\end{equation}
where
\begin{equation}
   G_+(k, \eta) =  \eta_0\frac{H^2}{k} \eta  e^{i k \eta} \, ,
\end{equation}
is the usual plus-time-ordered external propagator of a conformally-coupled scalar in the in-in formalism.
Using three properties~\eqref{eq:appA derivatives}, \eqref{eq:appA convolution}, and~\eqref{eq:appA G+} of the Mellin transform
\begin{align}
   \mcM(-\eta \partial_\eta f)(s) &= 2s f(s) \, ,\\
   \mcM(f_1 \dots f_n)(s) &= \int 2^n[\dd s]_n \deltabr(2s - 2s_T) f_1(s_1) \dots f_n(s_n) \label{eq:convolution} \, ,\\
   G_{+}(k, s) &\ceq \mcM(G_+(k,\eta))=   \frac{-\eta_0 H^2}{2} e^{- i \pi \left( s + \frac{1}{2} \right)} \Gamma\left( 2s + 1 \right) k^{-2s -2} \,.
\end{align}
where $\deltabr = 2\pi i \delta$ is the Dirac delta, $[\dd s]_n\ceq [\dd s_1]\ldots [\dd s_n]$, and $s_T = s_1 + \dots + s_n$. 
These results may be found in \cite{Paris_Kaminski_2001, Sleight:2019mgd}.
Specializing~\eqref{eq:convolution} to $2s=-3$ to evaluate the required time integral in~\eqref{timeinte}, we can attempt to calculate $B_{4s}$ with a formal sum $B_{4s}^\EFT$ of Mellin integrals
\begin{equation}
\begin{aligned}
   B_{4s}^\EFT = H^{-4} \, 2 \, \Re \sum_{n=0}^\infty 
   \int & [\dd^4 s] 2\deltabr(-3-2s_T) G_+(k_1, s_1)\dots G_+(k_4, s_4) \\
   \times & \frac{i\lambda^2}{M^2} \left[ -\frac{H^2}{M^2} \left( (2s_{34})^2 + 3 (2s_{34}) \right) \right]^n + (k_1,\,k_2\leftrightarrow k_3, \,k_4) \, .
   \label{eq:Mellin formal sum}
\end{aligned}
\end{equation}
The sum in $n$ still diverges, as discussed in the previous section. However, we can study the integrands themselves: if we formally bring the sum inside the integral, we find a series in Mellin space with coefficients $x^n = \left[ -\frac{H^2}{M^2} \left( (2s_{34})^2 + 3 (2s_{34}) \right) \right]^n$.
This series converges absolutely as long as $\abs{(2s_{34}^2) + 3(2s_{34}) } < M^2/H^2$, and corresponds to the Taylor expansion of $(1+x)^{-1}$.
We can blame the asymptotic nature of the derivative expansion of $B_{4s}^\EFT$ on the failure of this series to converge for all values taken by the integration variables $s_i$.
We can sum the series inside its radius of convergence and then perform an analytic continuation to obtain an integrand that is defined along the whole contour of integration. This produces another way to give meaning to the non-convergent sum of all EFT correlators:
\begin{align}
   B_{4s}^\EFT &= \frac{2}{H^4} \, \Re 
   \int [\dd^4 s] 2\deltabr(-3-2s_T) G_+(k_1, s_1)\dots G_+(k_4, s_4) \nonumber \\
   & \qquad  \times \frac{i\lambda^2}{M^2 + H^2\left( (2s_{34})^2 + 3(2s_{34}) \right)} + (k_1,\,k_2\leftrightarrow k_3, \,k_4)\, .
\end{align}
We can now perform the integrals over the Mellin variables: 
\begin{align}
   B_{4s}^\EFT
   &=  \left( -\eta_0 H \right)^4  e_4^{-\frac{1}{2}} 
   \int [\dd s]_{234} \, \Gamma\left( 2s_2 - \frac{1}{2} \right) \Gamma\left( 2s_3 - \frac{1}{2} \right) \Gamma\left( 2s_4 - \frac{1}{2} \right) \Gamma\left( \frac{5}{2} - 2 s_{234} \right)  \nonumber \\
   & \quad k_1^{-3} \left( \frac{k_1}{k_2} \right)^{2s_2} \left( \frac{k_1}{k_3} \right)^{2s_3}\left( \frac{k_1}{k_4}\right)^{2s_4} \frac{\lambda^2}{M^2 + H^2\left[ (2s_{34}-3)^2 + 3(2s_{34} - 3) \right]} + (k_1,\,k_2\leftrightarrow k_3, \,k_4) \, , \label{eq:B4s EFT integral}
\end{align}
where the integral over $s_1$ has been performed using the Dirac delta, and the integration variables have been shifted to match onto standard integrals. Here, $e_4 \ceq k_1 k_2 k_3 k_4$.
Next, we change variables to $s_\pm$ = $s_3 \pm s_4$.
The integrals in $s_2$ and $s_-$ are of the common form
\begin{equation}
   \int [\dd s] z^{2s} \Gamma(a+2s) \Gamma(b-2s) = \frac{1}{2} z^b \Gamma(a+b) (1+z)^{-a-b} \, .
   \label{eq:common integral}
\end{equation}
After performing those integrals, we decompose the rational function into partial fractions, then convert the partial fractions into gamma functions.
The resulting integral is
\begin{align}
   B_{4s}^\EFT &= \left( -\eta_0 H \right)^4 \left( 16  e_4 \right)^{-\frac{1}{2}} 
   \int [\dd s_+] \, \Gamma\left( 2s_+ - 1 \right)  \Gamma\left( 2 - 2s_+ \right)  k_1^{-3}  \left( \frac{k_1 k_4}{k_2 k_3} \right)^{\frac{1}{2}} \left( \frac{k_{12}}{k_1} \right)^{-2}     \nonumber \\
   & \quad   \left( \frac{k_{34}}{k_4} \right)  \left( \frac{k_{12}}{ k_{34}} \right)^{2s_+}  \frac{\lambda^2}{M^2 + H^2\left[ (2s_+-3)^2 + 3(2s_+ - 3) \right]} + (k_1,\,k_2\leftrightarrow k_3, \,k_4) \\
   &= \left( -\eta_0 H \right)^4 \left( 16 e_4 \right)^{-\frac{1}{2}}
   \int [\dd s_+] \, \Gamma\left( 2s_+ - 1 \right)  \Gamma\left( 2 - 2s_+ \right)  k_1^{-3}  \left( \frac{k_1 k_4}{k_2 k_3} \right)^{\frac{1}{2}} \left( \frac{k_{12}}{k_1} \right)^{-2}     \nonumber \\
   & \quad   \left( \frac{k_{34}}{k_4} \right)  \left( \frac{k_{12}}{ k_{34}} \right)^{2s_+}  \frac{\lambda^2}{2\mu H^2}  \left[ i\, \frac{\Gamma\left(2s_+-3 + \frac{3}{2}+i \mu\right)}{\Gamma\left( 2s_+-2 + \frac{3}{2} + i \mu \right)}  - i \, \frac{\Gamma\left(2s_+-3+ \frac{3}{2} - i \mu\right)}{\Gamma\left( 2s_+-2+ \frac{3}{2} - i \mu\right)}\right] + (k_1,\,k_2\leftrightarrow k_3, \,k_4)\, .
\end{align}
To perform the final integral in $s_+$, we must decide how to approach the new poles at $s_+ = -\frac{3}{2} \pm i\mu$ introduced by the resummation.


\subsection{Contour ambiguity}

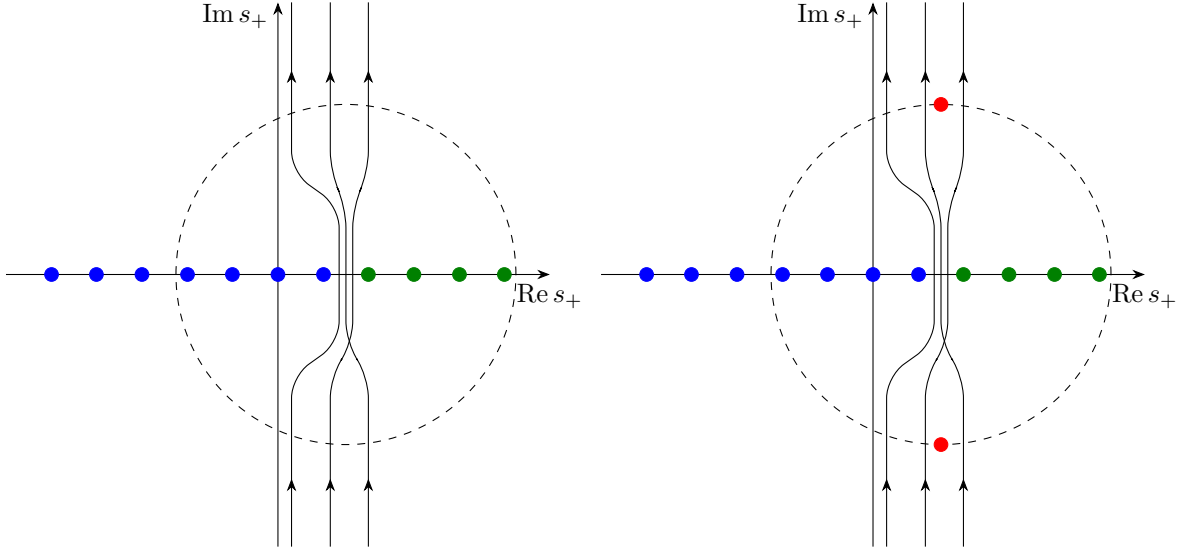
\begin{figure}[h]
   \centering
\begin{center}
   \begin{tikzpicture}[>=Stealth, scale=.9]
   \draw [->] (-4,0) -- (4,0);
   \draw [->] (0,-4) -- (0,4);
   \node[anchor=north] at (4,0) {$\Re s_+$};
   \node[anchor=east] at (0,3.8) {$\Im s_+$};

   \draw[dashed] (1,0) circle [radius=2.5];

   \filldraw[blue] (.67,0) circle [radius=.1];
   \filldraw[blue] (0,0) circle [radius=.1];
   \filldraw[blue] (-.67,0) circle [radius=.1];
   \filldraw[blue] (-1.33,0) circle [radius=.1];
   \filldraw[blue] (-2,0) circle [radius=.1];
   \filldraw[blue] (-2.67,0) circle [radius=.1];
   \filldraw[blue] (-3.33,0) circle [radius=.1];

   \filldraw[green!50!black] (1.33,0) circle [radius=.1];
   \filldraw[green!50!black] (2,0) circle [radius=.1];
   \filldraw[green!50!black] (2.67,0) circle [radius=.1];
   \filldraw[green!50!black] (3.33,0) circle [radius=.1];


   \draw[rounded corners=8pt] (.2, -4) -- (.2, -1.5) -- (.9, -1) -- (.9, 1) -- (.2, 1.5) -- (.2, 4);
   \draw[rounded corners=8pt] (1.33, -4) -- (1.33, -1.5) -- (1., -1) -- (1., 1) -- (.77, 1.5) -- (.77, 4);
   \draw[rounded corners=8pt] (.77, -4) -- (.77, -1.5) -- (1.1, -1) -- (1.1, 1) -- (1.33, 1.5) -- (1.33, 4);

   \path [tips,->] (.2, -4) to (.2, -3);
   \path [tips,->] (1.33, -4) to (1.33, -3);
   \path [tips,->] (.77, -4) to (.77, -3);

   \path [tips,->] (.2, 2) to (.2, 3);
   \path [tips,->] (1.33, 2) to (1.33, 3);
   \path [tips,->] (.77, 2) to (.77, 3);

\end{tikzpicture}
\begin{tikzpicture}[>=Stealth, scale=.9]
   \draw [->] (-4,0) -- (4,0);
   \draw [->] (0,-4) -- (0,4);
   \node[anchor=north] at (4,0) {$\Re s_+$};
   \node[anchor=east] at (0,3.8) {$\Im s_+$};

   \draw[dashed] (1,0) circle [radius=2.5];

   \filldraw[blue] (.67,0) circle [radius=.1];
   \filldraw[blue] (0,0) circle [radius=.1];
   \filldraw[blue] (-.67,0) circle [radius=.1];
   \filldraw[blue] (-1.33,0) circle [radius=.1];
   \filldraw[blue] (-2,0) circle [radius=.1];
   \filldraw[blue] (-2.67,0) circle [radius=.1];
   \filldraw[blue] (-3.33,0) circle [radius=.1];

   \filldraw[green!50!black] (1.33,0) circle [radius=.1];
   \filldraw[green!50!black] (2,0) circle [radius=.1];
   \filldraw[green!50!black] (2.67,0) circle [radius=.1];
   \filldraw[green!50!black] (3.33,0) circle [radius=.1];

   \filldraw[red] (1,2.5) circle [radius=.1];
   \filldraw[red] (1,-2.5) circle [radius=.1];

   \draw[rounded corners=8pt] (.2, -4) -- (.2, -1.5) -- (.9, -1) -- (.9, 1) -- (.2, 1.5) -- (.2, 4);
   \draw[rounded corners=8pt] (1.33, -4) -- (1.33, -1.5) -- (1., -1) -- (1., 1) -- (.77, 1.5) -- (.77, 4);
   \draw[rounded corners=8pt] (.77, -4) -- (.77, -1.5) -- (1.1, -1) -- (1.1, 1) -- (1.33, 1.5) -- (1.33, 4);

   \path [tips,->] (.2, -4) to (.2, -3);
   \path [tips,->] (1.33, -4) to (1.33, -3);
   \path [tips,->] (.77, -4) to (.77, -3);

   \path [tips,->] (.2, 2) to (.2, 3);
   \path [tips,->] (1.33, 2) to (1.33, 3);
   \path [tips,->] (.77, 2) to (.77, 3);

\end{tikzpicture}
\end{center}
   \caption{\textit{(Left) }For any term in the series~\eqref{eq:Mellin formal sum} of integrals, any complex linear combination  of the contours in the figure gives the same value to the integral, as long as the coefficients sum to $1$. The region of convergence of the series of Mellin integrands is illustrated by the dashed circle. \textit{(Right) }New poles are produced by summing the series of Mellin integrands. Depending on the coefficients of the contours on the left, the red poles' residues can be added to the value of the resummed integral.}
   \label{fig:contour choices}
\end{figure} 
When the new poles in the $B_{4s}^\EFT$ integral are on the left of the integration contour, the integral has a standard expression in terms of a ${}_2 F_1$ hypergeometric function, and passing either pole on the right introduces an additional residue.
If we interpret the rational function as the Mellin-space propagator of the heavy field, we are in principle free to add positive- and negative-frequency solutions to the propagator to change its boundary conditions, while it remains a Green's function.
These correspond to arbitrary multiples of the residues of the new poles.
The situation is illustrated in Figure~\ref{fig:contour choices}.
For scattering amplitudes on flat spacetime, this ambiguity is fixed by imposing Feynman boundary conditions on the heavy field.
Here, without a definite boundary condition for the field $\chi$, we parameterize the additional complementary functions by complex numbers \hbox{$D_+$ and $D_-$.}
The result is 
\begin{align}
   B_{4s}^\EFT &= (-\eta_0)^4 \frac{H^2 \lambda^2}{16 \mu} \frac{1}{e_4 k_{34}}  
   \left[ \frac{i}{ 1 + 2 i \mu} \, \pFq{2}{1}{1,\frac{1}{2}+i\mu}{\frac{3}{2} + i \mu}{- \frac{k_{12}}{k_{34}}} + (i\rightarrow-i) \right.
 \nonumber \\
     & \quad  \left.
     + \, D_+ \, \frac{i}{2} \left( \left( \frac{k_{12}}{k_{34}} \right)^{-\frac{1}{2} - i \mu} \Gamma\left( \frac{1}{2} - i \mu \right) \Gamma\left( \frac{1}{2} + i \mu \right)    \right) + D_- \, (i\rightarrow-i)\right] + (k_{12} \leftrightarrow k_{34} ) \\
     &= \frac{H^2 \lambda^2 \eta_0^4}{16\mu} \frac{1}{k_1 k_2 k_3 k_4 } \left[ 
        {-} \,\Re\left( I^{(2)}_{+ -}(k_{12}, k_{34}) + I^{(2)}_{+ -}(k_{34}, k_{12} )\right)
           \right. \label{eq:B4s EFT final}  \\ & \quad \left. 
              +\,i e^{-\pi\mu} D_+ \left(   I_+^{(1)}(k_{12}) \,  I_+^{(1)}(k_{34})^* + I_+^{(1)}(k_{12})^* \,  I_+^{(1)}(k_{34}) \right) 
           \right. \nonumber \\ & \quad \left.
           + \, i e^{\pi\mu} D_- \left(I_-^{(1)}(k_{12}) \,  I_-^{(1)}(k_{34})^* + I_-^{(1)}(k_{12})^* \,  I_-^{(1)}(k_{34})  \right)     \right] \, , \nonumber
\end{align}
in terms of the functions $I^{(1)}_\pm$ and $I^{(2)}_{\pm\pm}$ defined in~\eqref{eq:int1} and~\eqref{eq:int2}.
As $I_-^* = ie^{-\pi\mu}I_+$, defining 
\begin{align}
D\ceq D_+ + D_-      \,,
\end{align}
it is possible to write the trispectrum as
\begin{equation}
\begin{aligned}
    B_{4s}^\EFT  &= \frac{H^2 \lambda^2 \eta_0^4}{16\mu} \frac{1}{k_1 k_2 k_3 k_4 } \left[ 
        {-} \,\Re\left( I^{(2)}_{+ -}(k_{12}, k_{34}) + I^{(2)}_{+ -}(k_{34}, k_{12} )\right)
           \right. \\ & \quad \left. 
              +\,i e^{-\pi\mu} D \left(   I_+^{(1)}(k_{12}) \,  I_+^{(1)}(k_{34})^* + I_+^{(1)}(k_{12})^* \,  I_+^{(1)}(k_{34}) \right)  \right]
            \, , \label{eq:B4s EFT finalfinal}
\end{aligned}
\end{equation}
Unitarity requires that the trispectrum be real, from which it follows that  $D$ must be purely imaginary.
Comparing with the factorized~\eqref{eq:in-in trisp fact} and non-factorized~\eqref{eq:in-in trisp non-fact} contributions to the trispectrum in the in-in formalism, we see that under the identification 
\begin{align}
   2 i e^{-\pi \mu} D &= e^{-2\pi\mu}\abs{A}^2+ \abs{B}^2 \, ,
\end{align} 
we recover part of the factorized trispectrum.
However, no choice of the $D_\pm$s will reproduce additional parts of the full trispectrum, since these involve terms of the form $I^{(1)}_+I^{(1)}_+$ and $I^{(1)}_+I^{(1)*}_-$, which have a different $k$ dependence to the terms that appear here.
Here, the EFT is unable to reproduce aspects of the UV physics that depend on the phases of the $A$ and $B$ coefficients.
\subsection{Comparison with wavefunction resummation}
It is interesting to compare this correlator with the resummation of Section~\ref{sec:zero cs}.
The EFT wavefunction $\psi_{4s}^\EFT$ in~\eqref{eq:psi4 EFT} contains no free parameters and has the same set of $I$ functions as~\eqref{eq:B4s EFT final}.
In the Mellin-space resummation for the correlator, some exponentially small terms in $\mu$ are recovered, and there is more freedom to account for the initial state of the heavy field, but other exponentially small terms in $\mu$ are still missing from the low-energy theory in the Mellin-space resummation.
The boundary conditions of the wavefunction additionally allow the EFT for the wavefunction to be augmented with two free functions of momentum in the final correlator via $\psi_{4s}^\text{BC}$, as discussed in Appendix~\ref{app:psiBC}.
The wavefunction resummation can then reproduce all terms from the UV theory, at the cost of putting in their exact functional forms by hand.
The Mellin-space resummation can match more parts of the UV correlator than the EFT wavefunction resummation before it is augmented with boundary terms, but the wavefunction formalism with boundary terms is able to match the full correlator of the UV theory.

We might hope that the Mellin-space resummation would work also for a general massive scalar or, more simply, the non-dynamical massive field of Section~\ref{sec:NonDynField}. 
Heuristically, an operator with $n$ spatial derivatives will then give rise to Mellin-Barnes integrands involving gamma functions that look like
\begin{equation}
   \left( \frac{H^2}{M^2} \right)^n \Gamma(\alpha n + 2s + \beta) \Gamma(-2s) \left( \{k\} \right)^s \, ,
\end{equation}
for some $\alpha$ and $\beta$.
The problem is that, by the ratio test, a series with these terms has zero radius of convergence in the Mellin variable $s$. The procedure of simply summing a \textit{convergent} series in Mellin space followed by analytic continuation, which is what we were able to do in this section, fails. One could attempt Borel or Mittag-Leffler resummation of the Mellin integrands, but we will not explore this direction further in the present work.


\section{Conclusion and discussion}

In this work, we have studied the EFT expansion resulting from integrating out a massive field coupled to a light field in de Sitter. Focusing on two specific toy models, we have analyzed different ways to resum the otherwise divergent EFT contributions and we examined to what extent the resummation reproduces the UV results. We found that the EFT correctly resums to the full result when the massive field is non-dynamical, but more generally the resummation misses nonlocal terms which are exponentially suppressed in the heavy field mass for the Bunch-Davies state.

The starting point for our investigation was the fact that in the time-dependent backgrounds often encountered in cosmology, such as for example de Sitter or accelerated FLRW spacetimes, the contribution of operators of high-dimension $\Delta$ grows factorially in $\Delta$ at fixed couplings, hence contradicting the naive expectation from flat spacetime. This problem occurs both for operators with a large field multiplicity and for operators with a large number of derivatives. In the presence of sufficiently small couplings, the first few neglected high-dimension operators in perturbation theory do, indeed, give small contributions to observables because the factorial enhancement in $\Delta$ is still modest for not-too-large $\Delta$'s. In light of this, the approach in most of the literature so far has been to simply proceed as we usually do in flat spacetime and neglect all high-dimension operators beyond the leading few. While this may be a viable phenomenological approach, it comes with some drawbacks. First, conceptually, one would like to possess some mathematical framework to make sense of neglecting infinitely many operators whose contribution is large. More precisely, one would like to learn the precise rules for setting up a \textit{consistent power counting scheme in cosmological EFTs}, which takes into account the specificities of the time-dependent background. Second, one would like to learn how to deal with EFT power counting schemes with \textit{multiple competing scales}. If an infinite family of high-dimension operators is suppressed by a much larger scale than that of a second infinite family, one would expect that, at a given precision, fewer operators of the first family need to be included compared to the second family. But if both families include infinitely many operators with arbitrary large contributions to observables this expectation seems to rest on shaky mathematical foundations. Third, one may want to learn to what extent building EFTs from integrating out sectors that are separated in energy is a good idea in time-dependent backgrounds where energy is not a conserved quantum number. 

Our work could be extended in a variety of different directions. For example, it would be interesting to study more realistic cases, like the exchange of a scalar field with unit speed of sound. Here the full result is known \cite{Arkani-Hamed:2018kmz} but it is much more complicated. Also, it would be nice to investigate how the set of radiatively-stable EFT power-counting schemes differs in flat and time-dependent backgrounds.


\paragraph{Acknowledgements} We would like to thank Calvin Chen, Jordan Cotler, Laura Engelbrecht, Sadra Jazayeri, Mang Hei Gordon Lee, Scott Melville, Ryo Namba, Guilherme Pimentel, Andrea Sanfilippo, and Yoav Zigdon for useful discussions. E.P. has been supported in part by the research program VIDI with Project No. 680-47-535, which is (partly) financed by the Netherlands Organisation for Scientific Research (NWO). C.M.~is supported by Science and Technology Facilities Council (STFC) training grant ST/W507350/1. This work has been partially supported by STFC consolidated grant ST/T000694/1 and ST/X000664/1 and by the EPSRC New Horizon grant EP/V017268/1. H.G. is supported by a Postdoctoral Fellowship at National Taiwan University funded by the National Science and Technology Council (NSTC) 113-2811-M-002-073. The work of C.D.P. is supported by the Scuola Normale Superiore.

\appendix

\section{Mittag-Leffler Summation}\label{app:ML}
In this Appendix we provide a brief overview of the Mittage-Leffler summation method based on~\cite{hardy2024divergent} which we used in Section~\ref{sec:NonDynField} to resum the asymptotic series produced by the EFT. This method is a generalization of Borel summation which is better adapted to our purposes. It can also be used to ensure the convergence of the infinite sum over $n$ in~\eqref{eq:flipsum} at all values of $a$ which recovers the same result as in~\eqref{eq:psi4 EFT}. We first define our function as a formal power series,
\begin{align}
    f(z)=\sum_{n=0}^\infty a_n z^n.
\end{align}
Taking this sum as a starting point we define
\begin{align}
    a(x)=\sum_{n=0}^\infty \frac{a_n z^n}{\Gamma(1+\alpha n)}x^{\alpha n}.
\end{align}
Integrating this new function term-by-term gives
\begin{align}
    \int_{0}^\infty e^{-x} a(x) \dd x=\sum_{n=0}^\infty \frac{a_n z^n}{\Gamma(1+\alpha n)}\int_{0}^\infty x^{\alpha n} e^{-x} \dd x=\sum_{n=0}^\infty a_n z^n.
\end{align}
From this relationship we define the sum
\begin{align}
    s(B,\alpha)\ceq \int_{0}^\infty \dd xe^{-x}\sum_{n=0}^\infty  \frac{a_n z^n x^{\alpha n}}{\Gamma(1+\alpha n)}.
\end{align}
Then we take this sum to define the resummed series
\begin{align}
    f(z)=s(B,\alpha).
\end{align}
In this way we can make use of the improved convergence properties of the sum
\begin{align}
    \sum_{n=0}^\infty a_n \frac{a_nz^n x^{\alpha n}}{\Gamma(1+\alpha n)} \, ,
\end{align}
compared to our original sum which, in the cases considered here, is enough to assign a finite value to our EFT series.

\section{Boundary Wavefunction Coefficient}\label{app:psiBC}
In this Appendix we demonstrate how we can choose a particular complementary function to add to the infinite tower of local differential operators to exactly recover the true correlator and wavefunction coefficient. The total contribution to the action from the boundary is, cf.~\eqref{334},
\begin{align}
    S_{\text{BD}}=\int_\bfx \left[\frac{1}{2}a^2 \chi \partial_\eta \chi+\lambda a^2\chi_{\text{CF}}\partial_\eta \chi -\lambda a^2\chi \partial_\eta \chi_{\text{CF}}\right]_{-\infty}^{\eta_0}\,.
\end{align}

To begin with, we parameterize the complementary function just as we did for the exact solution,
\begin{align}\label{eq:XCF}
    \chi_{\text{CF}}=\frac{i(-\eta)^{3/2}}{2\mu H^2\eta_0^2}\left(\mathcal{A}(k)\left(\frac{\eta}{\eta_*}\right)^{i\mu}+\mathcal{B}(k)\left(\frac{\eta}{\eta_*}\right)^{-i\mu}\right)\bar{\varphi}^2\,.
\end{align}
Where the boundary value of the conformally coupled field, $\bar{\varphi}$, is necessary for matching to wavefunction coefficients.
Likewise, the scaling with $\eta_0$ may seem adhoc but, as we will see, it is required to match the scaling of the exact solution. 
Note that we are required to introduce some momentum dependence on the coefficients $\mathcal{A}$ and $\mathcal{B}$, despite the $\chi$ field itself not having any such dependence. We can understand this as coming from the momentum of the external fields. On this boundary the contribution from the infinite sum can be evaluated similarly to the integral by an exchange of the sums,
\begin{align}\label{eq:chisol2}\nonumber
    \chi&=\frac{\lambda(-\eta)^2}{H^2\mu^2\eta_0^2}\sum_{b=0}^{\infty}\sum_{a=0}^b\sum_{n=0}^\infty \left(-\frac{1}{\mu^2}\right)^{n}e^{ik_{12}\eta}(-1)^{a+b}\binom{b}{a}\left(a+\frac{1}{2}\right)^{2n}\frac{(ik_{12}\eta)^b}{b!}\bar{\varphi}^2+\lambda\chi_{\text{CF}}\\&=\frac{i\lambda\eta^2}{2\mu H^2 \eta_0^2}(-ik_{12}\eta)^{-\frac{1}{2}-i\mu}\left((-ik_{12}\eta)^{2i\mu}\gamma\left(\frac{1}{2}-i\mu,-ik_{12}\eta\right)-\gamma\left(\frac{1}{2}+i\mu,-ik_{12}\eta\right)\right)\bar{\varphi}^2+\lambda\chi_{\text{CF}}\,.
\end{align}
Here we have introduced the lower incomplete gamma function,
\begin{align}
    \gamma(s,x)=\int_0^x t^{s-1} e^{-t} \dd t\,.
\end{align}
We could have similarly computed the solution to the equations of motion in the full theory by performing a time integral over the Green's function. This exact solution will necessarily satisfy the boundary conditions imposed on the Green's function that it vanishes in both the infinite past and future,
\begin{align}\nonumber\label{eq:ChiCF}
    \chi_{\text{exact}}=& \,\frac{i\lambda \eta^2}{2\mu H^2\eta_0^2}(-ik_{12}\eta)^{-1/2-i\mu}\left((-ik_{12}\eta)^{2i\mu}\gamma\left(\frac{1}{2}-i\mu,-ik_{12}\eta\right)-\gamma\left(\frac{1}{2}+i\mu,-ik_{12}\eta\right)\right)\bar{\varphi}^2\\
    &+i\frac{\lambda(-\eta)^{3/2} }{2\mu H^2\eta_0^2}\frac{A^*I^{(1)}_-(k_{12})+B^* I^{(1)}_+(k_{12})}{A^*(\eta_0/\eta_*)^{-i\mu}+B^*(\eta_0/\eta_*)^{i\mu}}\left(\left(\frac{\eta}{\eta_0}\right)^{i\mu}-\left(\frac{\eta}{\eta_0}\right)^{-i\mu}\right) { \bar{\varphi}^2} \, .
\end{align}
We can see that this first line is precisely the result of summing over all the EFT contributions and so the second line must equal $\chi_{\text{CF}}$ in~\eqref{eq:XCF}. We can enforce this matching by setting
\begin{align}
    \mathcal{A}(k_{12})&=\frac{A^*I^{(1)}_-(k_{12})+B^* I^{(1)}_+(k_{12})}{A^*(\eta_0/\eta_*)^{-i\mu}+B^*(\eta_0/\eta_*)^{i\mu}}\left(\frac{\eta_0}{\eta_*}\right)^{{-}i\mu}=-\mathcal{B}(k_{12})\left(\frac{\eta_0}{\eta_*}\right)^{{-}2i\mu}\,.
\end{align}
We can also see, most clearly in~\eqref{eq:ChiCF}, that this solution has $ \chi_{\text{CF}}(\eta_0)=0$. This is compatible with the results presented in~\cite{salcedo2023analytic} as the infinite series has no contribution at $\mathcal{O}(\eta_0^{-2})$ in the limit that we send $\eta_0\rightarrow 0$. Therefore, the future boundary term will vanish. However, there is still a contribution from the infinite past.
In performing this matching we have set the two solutions to be identical and we must get the same answer. Explicitly evaluating this on the past boundary then gives
\begin{align}\label{eq:psiBD}
    \psi_{4s}^{\text{BD}}&=-\frac{\lambda^2}{4\mu H^6\eta_0^4}\Big(\mathcal{A}(k_{12})I_+^{(1)}(k_{34})+\mathcal{B}(k_{12})I_-^{(1)}(k_{34})\Big)\bar{\varphi}^4+(k_{12}\leftrightarrow k_{34})\,.
\end{align}
Adding this to the previously calculated EFT contribution gives the full wavefunction,
\begin{align}\nonumber
    \psi^{\text{EFT}}_{4s}+\psi^{\text{BD}}_{4s} =&\,-\frac{ \lambda^2}{2\mu H^6\eta_0^4}\Bigg[I^{(2)}_{+-}(k_{12},k_{34})+I^{(2)}_{+-}(k_{34},k_{12})+\frac{1}{A^*(\eta_0/\eta_*)^{-i\mu}+B^*(\eta_0/\eta_*)^{i\mu}}\left(\frac{\eta_0}{\eta_*}\right)^{i\mu}\\\nonumber
    &\times\Big[-2B^*\left(I^{(1)}_+(k_{12})I^{(1)}_-(k_{34})+I^{(1)}_-(k_{12})I^{(1)}_+(k_{34})\right)\\
    &\qquad-A^*I^{(1)}_-(k_{12})I^{(1)}_-(k_{34})+B^*I^{(1)}_+(k_{12})I^{(1)}_+(k_{34})\Big]\Bigg]\,.
\end{align}
This agrees with the exact result that we calculated previously in Section~\ref{sec:cs0exact}.
However, as mentioned in the main text, this procedure is not well defined within the EFT as it requires adding in a contribution, $\chi_{\text{CF}}$, that can only be computed using the full theory. 

Alternatively, in~\cite{Green:2024cmx}, to match the correlator, they take the late time limit of this boundary term to be 
\begin{align}\nonumber
    \lim_{\eta\rightarrow \eta_0}\chi&=\frac{\psi_3+\psi_3^*}{\psi_2^{\chi}+(\psi_2^{\chi})^*}\bar{\varphi}^2= \frac{i\lambda}{2\mu H^2(-\eta_0)^{1/2}}\left(\frac{\eta_0}{\eta_*}\right)^{-i\mu} \\\nonumber&\times\Bigg[\bigg((1+|B|^2(1-ie^{\pi\mu}) I^{(1)}_- (k) +AB^* (1-ie^{-\pi\mu}) I^{(1)}_+ (k)\bigg)\left(\frac{\eta_0}{\eta_*}\right)^{2i\mu} \\& \qquad+\bigg( BA^*(1-ie^{\pi\mu}) I^{(1)}_- (k) + (|B|^2 (1-i e^{-\pi\mu})-i e^{-\pi\mu})I_+^{(1)}(k)\bigg) \Bigg]\bar{\varphi}^2\,.
\end{align}
As we saw in the previous discussion, the infinite sum coming from the EFT terms in $\chi$ vanishes on the future boundary. To match this to the $\eta\to\eta_0$ limit of the complementary function~\eqref{eq:XCF} we will expand our coefficients as
\begin{align}
    \mathcal{A}(k)=\mathcal{A}_-I_-^{(1)}(k)+\mathcal{A}_+I_+^{(1)}(k)\,,\qquad\mathcal{B}(k)=\mathcal{B}_-I_-^{(1)}(k)+\mathcal{B}_+I_+^{(1)}(k)\,.
\end{align}
Then, we find that these coefficients must be given by
\begin{equation}
\begin{aligned}\label{eq:BCs}
    \mathcal{A}_-&=1+|B|^2(1-ie^{\pi\mu})\,,&
    \mathcal{B}_+&=-ie^{-\pi\mu}(1+|B|^2(1+ie^{\pi\mu}))\,,\\
    \mathcal{A}_+&=AB^*(1-ie^{-\pi\mu})\,,&
    \mathcal{B}_-&=A^*B(1-ie^{\pi\mu})\,.
\end{aligned}
\end{equation}
Imposing this boundary condition we find that the contribution to $\psi_{4s}^{\text{BD}}$ from the late time limit vanishes at $\mathcal{O}(\eta_0^{-4})$ so we once again only need the early time boundary. Similarly, the boundary term $[a^2\chi\partial_\eta\chi]$ will vanish in the past due to our vacuum condition on the mode functions. Thus, the boundary contribution is given, as before, by~\eqref{eq:psiBD}. 
As there is no three point interaction in this EFT the correlator is straightforwardly given by 
\begin{align}
    &\frac{2\Re\left(\psi^{\text{EFT}}_{4s}+\psi^{\text{BD}}_{4s} \right)}{2\Re\left(\psi_2^{\varphi}\right)^4}=-\frac{\lambda^2H^2\eta_0^4}{16\mu k_1k_2k_3k_4}\left[\Re(I_{+-}^{(2)}(k_{12},k_{34})+I_{+-}^{(2)}(k_{34},k_{12}))\right.\\
    &\left.\,+\,\Re\left[(\mathcal{A}_+-\mathcal{B}_-^*e^{-2\pi\mu}) I_+^{(1)}(k_{12})I_+^{(1)}(k_{34})\right]+e^{-\pi\mu}\Im(\mathcal{A}_-+\mathcal{B}_+)\,\Re\left(I^{(1)}_+(k_{12})I^{(1)}_+(k_{34})^*\right)\right] \, . \nonumber
\end{align}
The exact correlator~\eqref{eq:B4correl} can be expressed as
\begin{align}\label{eq:B4exact}
    B_{4s} = & \, -\frac{\lambda^2H^2\eta_0^4}{16\mu k_1k_2k_3k_4} \Bigg[   \Re\left(I^{(2)}_{+-} (k_{12},k_{34})+I^{(2)}_{+-} (k_{34},k_{12})\right){ +}\,\Re\left(AB^*(1-ie^{-\pi\mu})^2\, I^{(1)}_{+}(k_{12})I^{(1)}_{+}(k_{34})\right) \nonumber
    \\& \, -(e^{-2\pi\mu}+(e^{-2\pi\mu}+1)|B|^2)\,\Re\left(I_+^{(1)}(k_{12})I_+^{(1)}(k_{34})^*\right)\Bigg] \, .
\end{align}
Equality between these two is thus ensured if
\begin{equation}
\begin{aligned}
    -e^{-\pi\mu} -(e^{\pi\mu}+e^{-\pi\mu}) |B|^2=\Im(\mathcal{A}_-+\mathcal{B}_+),\\
    AB^* \,(1-ie^{-\pi\mu} )^2=\mathcal{A}_+-\mathcal{B}_-^*e^{-2\pi\mu}.
\end{aligned}
\end{equation}
Which is satisfied by the relationships in~\eqref{eq:BCs}. Therefore, we can recover both the correlator and the wavefunction coefficients by an appropriate choice of $\mathcal{A}$ and $\mathcal{B}$ in the complementary function. 

\section{Mellin transform and Feynman rules} \label{app:Mellin}
The properties of the Mellin transform are described in \cite{Paris_Kaminski_2001}, and particularly useful properties are summarized here.
Following  conventions\footnote{In \cite{Paris_Kaminski_2001}, the variable $z=2s$ is used instead. It is easy to pass between conventions, but we will use the one which simplifies comparison with previous work in cosmology.} in \cite{Sleight:2019mgd}, the Mellin transform of a function $f$ is defined as
\begin{equation}
   \mcM(f)(s) = f(s) \ceq \int_{-\infty}^{0} \frac{\dd \eta}{-\eta} \, (-\eta)^{2s} f(\eta)\,, 
\end{equation}
when the integral converges.
Changing variables to $2s=i\omega$ and $-\eta=H^{-1} e^{-t}$, the Mellin transform becomes a standard Fourier transform: this sets expectations for its properties as an integral transform.
If the integral converges absolutely for
\begin{equation}
   a < 2\Re s < b \, ,
\end{equation}
which requires that $f(\eta)$ decays faster than $(-\eta)^{-a}$ at $\eta=0$ and $(-\eta)^{-b}$ in the infinite past, then we refer to this region as the fundamental strip or the strip of analyticity in the complex $s$ plane, and $f(s)$ is analytic there. We therefore have that
\begin{equation}
   f(\eta) = \frac{1}{2\pi i} \int_{c-i\infty}^{c+i\infty}  2\dd s \, (-\eta)^{-2s}f(s) \eqc \int  2[\dd s] (-\eta)^{-2s}f(s) \, ,
\end{equation}
where the vertical contour lies in the fundamental strip, and the notation $[\dd s]$ means that the integral should be normalized by $1/2\pi i$ and should pass along the appropriate Mellin inversion contour.
When $n$ such integrals are to be performed, the notation $[\dd s]_n$ is used.

\subsection{Properties of the Mellin transform}
We describe the Mellin-space analogues of some familiar, useful properties of the Fourier transform.
These will allow us to translate the Feynman rules to Mellin space in much the same way as to Fourier space.

\paragraph{Derivatives}
Conformal time derivatives are diagonal in the power law basis:
\begin{equation}
   \mcM(-\eta \partial_\eta f)(s) = 2s f(s)\,. \label{eq:appA derivatives}
\end{equation}
in the fundamental strip.

\paragraph{Shifts}
Multiplication by $-\eta$ corresponds to a shift in s:
\begin{equation}
   \mcM\left((-\eta)^{r} f\right)(s) = f\left( s + \frac{r}{2} \right) \, ,
\end{equation}
which also has the effect of shifting the fundamental strip.

\paragraph{Generalized functions}
It is also useful to have the Mellin transform of a pure power law and, for completeness, the Dirac delta, which follow from the Fourier transforms of a complex exponential and the Dirac delta:
\begin{align}
   (-\eta)^r &= \int 2 [\dd s] (-\eta)^{-2s} \deltabr(2s + r)\,, \\
   (-\eta_0)^{2s} &= \int \dd \ln (-\eta) \, (-\eta)^{2s} \delta(\eta - \eta_0)\,.
\end{align}
where the bracketed Dirac delta comes with a factor of $2\pi i$, and the contour of integration formally passes through $2s=-r$ parallel to the real axis.
In each case, the forward or inverse transform will formally fail to converge.
The transform of a step function is also desirable:
\begin{equation}
   \int \dd \eta (-\eta)^{2s - 1} \theta(\eta - \eta_0) = \frac{(-\eta_0)^{2s}}{2s} \, .
\end{equation}
The integral converges only for $\Re s>0$, so the contour of integration passes to the right of this pole.
One can similarly derive a Sokhotski-Plemelj theorem if needed.

\paragraph{Integrals and products}
Integrals over all time are given by a particular value of the Mellin transform:
\begin{equation}
   \int_{-\infty}^{0} \dd \eta \, f(\eta) = f\left( s=\frac{1}{2} \right) \, ,
\end{equation}
which is trivial given the definition of the Mellin transform, but in combination with convolution, this property allows the elimination of time integrals. \\
There is a Parseval's theorem:
\begin{align}
   \int_{-\infty}^{0} \frac{\dd \eta}{(-\eta)} f(\eta) g(\eta) &= \int 2 [\dd s] f(s) g(-s)\,, \\
   \int_{-\infty}^{0} \dd \eta \, f(\eta) g(\eta) &= \int 2 [\dd s] f(s) g\left( \frac{1}{2} - s \right)\,,
\end{align}
where the first line is a change of variables in the familiar result for the Fourier transform, and the second line is a result of the shift property. \\
There is also a convolution theorem:
\begin{equation}
   \mcM(f_1  \dots f_n)(s) = \int 2^n[\dd s]_n \deltabr(2s - 2s_T) f_1(s_1) \dots f_n(s_n) \,.\label{eq:appA convolution}
\end{equation}
as long as the integrals converge, so
\begin{equation}
   \int_{-\infty}^{0} \dd \eta \, f_1(\eta) \cdot \dots f_n(\eta)= \int 2^n[\dd s]_n \deltabr(1 - 2s_T) f_1(s_1) \dots f_n(s_n)\,.
\end{equation}

\subsection{Feynman integrals}
Other authors have already used these properties of the Mellin transform to develop a diagrammatic Mellin-space approach to correlators \cite{Sleight:2021plv,Sleight:2021iix}. Here we describe an essentially equivalent formalism.
We will work in the in-in formalism.
Application to the wavefunction formalism is straightforward.
We will study a Feynman diagram with $V_{\pm}$ vertices, $E$ external lines, and $I$ internal lines forming $L$ loops.
The $i$th internal line will carry momentum $q_i$ between vertices $A_i$ and $B_i$, with $N^A_i$ and $N^B_i$ time derivatives, respectively.
The $e$th external line will carry momentum $k_e$ from vertex $E_e$ and has $N^E_e$ time derivatives.
The $v$th vertex has $D_v$ spatial derivatives, resulting in some invariant structure $\{\bfq_v\}^{D_v}$, and $N_v$ time derivatives appearing to the right of all spatial derivatives (for definiteness, but it is easy to adapt to other situations).
The individual momenta (external and internal) flowing into the vertex are $\bfq_v^{(i)}$.
This vertex is (anti-)time ordered according to the sign $\pm_v$, and has coupling $\lambda_v$.
The diagram takes the form
\begin{align}
   \mcD &= \left[\prod_{i \in I} \int \dbar^d \bfp_i \prod_{v\in V} \int \frac{\dd \eta_v}{(-H\eta_v)^{d+1}} (\pm_v i \lambda_v) (-H\eta_v)^{D_v} \{\bfq_v\}^v \deltabar\left( \sum_{i} \bfq_v^{(i)} \right) \right]  \\ 
   & \quad \times \prod_{i \in I} \left( -H\eta_{A_i} \partial_{\eta_{A_i}} \right)^{N^{A}_{i}} \left( -H\eta_{B_i} \partial_{\eta_{B_i}} \right)^{N^{B}_{i}} G_{\pm_{A_i} \pm_{B_i}}(q_i, \eta_{A_i}, \eta_{B_i}) \prod_{e \in E}  \left( -H\eta_{E_e} \partial_{\eta_{E_e}} \right)^{N^{E}_{e}} G_{\pm_{E_e}}(k_e, \eta_{E_e})\,. \nonumber
\end{align}
We can rewrite the time integrals in Mellin space, handling the products with the convolution theorem and generating Mellin variables for (half-)lines and $\delta$s at vertices
\begin{align}
   \mcD &= \left[ \prod_{i \in I} \int \dbar^d \bfp_i \prod_{i \in I} 2^2[\dd s^{I}_{A_i} \dd s^{I}_{B_i}] \prod_{e \in E} 2[\dd s^{E}_e] \right] \nonumber \\
   & \quad \times \prod_{v\in V} (\pm_v i \lambda_v) H^{-d-1+D_V + N_V} \{\bfq_v\}^v \deltabar\left( \sum_{i} \bfq_v^{(i)} \right) 2[\delta]\left( D_v - d - 2 s_{T}^{I_v} -2 s_{T}^{E_v} \right) \\ 
   & \quad \times \prod_{i \in I} (s_{A_i}^I)^{N^A_i} (s_{B_i}^I)^{N^{B}_i} G_{\pm_{A_i} \pm_{B_i}}(q_i, s^{I}_{A_i}, s^{I}_{B_i}) \prod_{e \in E} (s^{E}_e)^{N^{E}_e}G_{\pm_{E_e}}(k_e, s^{E}_e) \, . \nonumber
\end{align}
This expression will allow us to write down Feynman rules in Mellin space.
It appears that we have $2I + E - V = E + I + L - 1 = E + V + 2L - 2$ integrals over $s$ to perform.
On flat space, the Feynman propagator is monochromatic: it is only supported at $\omega^{I}_{A_i} = - \omega^{I}_{B_i}$.
On de Sitter, the step function in the Feynman propagator will bring an additional integral for each propagator.

We need to know the Mellin transforms of the various propagators.
These appear in \cite{Sleight:2019mgd}, with the Mellin transform of Hankel functions also given in \cite{Paris_Kaminski_2001}.
An $i\varepsilon$ prescription enforced by giving the momenta small imaginary parts, which ensures that the Mellin integrals converge on contours parallel to the imaginary axis.
The non-Feynman propagators are
\begin{align}
   G_{\pm}(k, s) &=  \frac{N_\nu}{2} e^{\mp i \pi \left( s + \frac{d}{4} -  \frac{\nu}{2} \right)} \Gamma\left( s + \frac{d}{4} + \frac{\nu}{2} \right) \Gamma\left( s + \frac{d}{4} - \frac{\nu}{2} \right) \left( \frac{k}{2} \right)^{-2s - \frac{d}{2} - \nu} \,, \\
   N_\nu &\ceq  (-\eta_0)^{\frac{d}{2} - \nu}H^{d-1} \frac{\Gamma(\nu)}{4\pi}\,, \\
   G_{\pm \mp}(q, s_A, s_B) &= \frac{1}{4} \frac{1}{4\pi} e^{\mp i \pi \left( s_A + \frac{d}{4} -  \frac{\nu}{2} \right)} \Gamma\left( s_A + \frac{d}{4} + \frac{\nu}{2} \right) \Gamma\left( s_A + \frac{d}{4} - \frac{\nu}{2} \right) \left( \frac{q}{2} \right)^{-2s_A - \frac{d}{2}} \nonumber \\
   & \quad
    e^{\pm i \pi \left( s_B + \frac{d}{4} -  \frac{\nu}{2} \right)} \Gamma\left( s_B + \frac{d}{4} + \frac{\nu}{2} \right) \Gamma\left( s_B + \frac{d}{4} - \frac{\nu}{2} \right) \left( \frac{q}{2} \right)^{-2s_B - \frac{d}{2}}\,.
\end{align}
where a dependence on the final time at which the correlator is evaluated has been omitted from $N_\nu$, corresponding to taking the leading component at late times
For conformally-coupled scalars in $d=3$, $\nu=1/2$ and the external propagator is
\begin{equation}
   G_{\pm}(k, s) =  2N_{1/2} \pi^{1/2} e^{\mp i \pi \left( s + \frac{1}{2} \right)} \Gamma\left( 2s + 1 \right) k^{-2s -2}  \, . \label{eq:appA G+}
\end{equation}
For the Feynman propagator, we need
\begin{align}
   (\mcM \theta(\eta_1 - \eta_2))(s_1, s_2) &= \int_{-\infty}^{0} \dd \eta_1 \int_{-\infty}^{\eta_1} \dd \eta_2 (-\eta_1)^{2s_1-1}(-\eta_2)^{2s_2-1} \nonumber \\
   &= \int_{-\infty}^{0} \dd \eta_1 (-\eta_1)^{2s_1+2s_2-1}\frac{-1}{2s_2 + \varepsilon} \\
   &= \frac{-1}{2s_2 + \varepsilon} [\delta](2s_1 + 2s_2)\,. \nonumber 
\end{align}
requiring $\Re s_2 < 0$ for convergence; the $\varepsilon > 0$ is a reminder that the pole should be on the right of the $s_2$ contour, arranging for this to happen if the contour is placed on the imaginary axis.
Now, using the convolution theorem and the factorized propagators, we obtain
\begin{align}
   G_{++}(q, s_A, s_B) &= \int 4[\dd s_{a_1} \dd s_{a_2}] \int 4[\dd s_{b_1} \dd s_{b_2}] \deltabr(2s_A - 2s_{a_1} - 2s_{a_2}) \deltabr(2s_B - 2s_{b_1} - 2s_{b_2}) \nonumber \\
   &\quad \times\frac{-1}{2s_{b_2} + \varepsilon} \deltabr(2s_{a_2} + 2s_{b_2}) G_{-+}(q, s_{a_1}, s_{b_1}) + (s_A \leftrightarrow s_B) \nonumber \\
   &= \int 4[\dd s_{a_1} \dd s_{b_q}] \int 2[\dd u] \deltabr(2s_A - 2s_{a_1} + 2u) \deltabr(2s_B - 2s_{b_1} - 2u) \\
   &\quad \times\frac{-1}{2u + \varepsilon} G_{-+}(q, s_{a_1}, s_{b_1}) + (s_A \leftrightarrow s_B) \nonumber\\
   &= \int 2[\dd u] \frac{1}{2u - \varepsilon} G_{-+}(q, s_{A}- u, s_{B} + u) + (s_A \leftrightarrow s_B) \, , \nonumber
\end{align}
and there may not be much hope of evaluating the integral for general $\nu$.
In any case, we will not need the Mellin-space Feynman propagator for the calculations in the main text, and it is only included here for completeness.

If we do not perform the Mellin integrals of the external lines, and for simplicity remove the external propagators themselves, we can obtain an amputated correlator in Mellin space with $E$ external lines $A(s_1\dots s_E, \pm_1\dots\pm_E)$.
This is because in calculating a correlator $\ev{\phi^n}$ of a field $\phi$ in the in-in formalism, each external propagator of a Feynman diagram must be labeled as either time-ordered $(+)$ or anti-time ordered $(-)$.
The contribution to the correlator from each possible assignment can be written as an integral of external propagators multiplying a function $A$ defined in Mellin space: schematically,
\begin{align}
   \ev{\phi^n}'= B_n &= \sum \text{all connected Feynman diagrams $\mcD$} \nonumber\\
   &= \sum_{ \pm_1\dots\pm_n } \sum \text{$\mcD$ such that the $i$th external line has a $G_{\pm_i}$ propagator} \\
&= \sum_{ \pm_1\dots\pm_n } \int [\dd^{n} s] G_{\pm_{v_1}} (k_1, s_1) \dots G_{\pm_{v_n}} (k_n, s_n) A\left( s_1, \dots, s_n; \pm_1\dots\pm_n \right)\,, \nonumber
\end{align}
where $A$ accounts for all possible diagram topologies compatible with the choice of $\pm$ propagators on each of the external legs.
The final line may be written as an integral in Mellin space thanks to the theorem \eqref{eq:appA convolution}, which relates integrals of products in conformal time to integrals in Mellin space.
Because of the similarity of this final set of integrals to the procedure of \cite{Melville:2024ove} to pass from an amputated correlator to an $S$-matrix element, or the procedure in \cite{Donath:2024utn} which gives correlators in terms of off-shell amplitudes, it is tempting to interpret these final integrals as turning an amputated correlator $A$ in Mellin space into an observable at the end of inflation.

The Feynman rules to calculate this $A$ in Mellin space can be given as a transformation of the conformal-time Feynman rules:
\begin{itemize}
   \item Label each external line and each half of each internal line with a Mellin $s$ variable.
   \item Amputate the external propagators and replace each internal propagator with the appropriate double Mellin integrand.
   \item A vertex from an interaction $\lambda \phi^n/n!$ at which external lines with Mellin variables $s^{E}_i$ and internal half-lines with variables $s^{I}_i$ carries a factor $\lambda\, 2[\delta](-d - 2 \sum_i s^{E}_i - 2 \sum_i s^{I}_{i})$, constraining the sum of Mellin variables meeting at that point.
   \item If there are $D_v$ spatial derivatives at that vertex, the Dirac delta is instead $2[\delta](D_v - d - 2 \sum_i s^{E}_i - 2 \sum_i s^{I}_{i})$. Spatial derivatives introduce factors of $i\bfk$ as usual.
   \item Each vertex also carries the usual momentum-conserving Dirac delta.
   \item Every $-\eta\partial_\eta$ becomes the Mellin $2s$ of its argument.
   \item Integrate over all internal momenta and Mellin variables
   \item Sum over all connected diagrams compatible with the $\pm$ time-orderings of the external legs.
   \item The corresponding contribution to the correlator at the end of inflation is found by re-introducing the external propagators and integrating over all of the external Mellin variables.
\end{itemize}
To pass from these $A$s to a correlator at the end of inflation, one takes $G_{\pm_1}(s_1)\dots G_{\pm_E}(s_E) A(s, \pm_1\dots\pm_E)$ and integrates over the free Mellin variables, then sums over all of the $\pm$ time orderings associated to the external legs.

When calculating the tree-level trispectrum in the EFT of Section~\ref{sec:EFT2}, only contact interactions contribute: all external legs of the Feynman diagrams are attached to the same vertex and so all must have the same $+$ or $-$ time ordering.
Therefore, only $A(s_1,\dots,s_4;++++) $ and $A(s_1,\dots,s_4;----)$ are nonzero and we define
\begin{align}
    A(s_1,\dots,s_4;++++) \eqc A^+(s_1\dots s_4)\,, \quad A(s_1,\dots,s_4;----) \eqc A^-(s_1\dots s_4)\,. 
\end{align}
The contribution to $A^+$ from the interaction
\begin{equation}
   \mcO_n \ceq \frac{\lambda^2}{M^2} \left[ -\frac{H^2}{M^2} \left( \eta\partial_\eta(\eta\partial_\eta) - 3 \eta\partial_\eta \right) \right]^n\,,
\end{equation}
is
\begin{equation}
   A^+_n(s_1\dots s_4) =2\deltabr(-3-s_T) \frac{i\lambda^2}{M^2} \left[ -\frac{H^2}{M^2} \left( (2s_{34})^2 - 3 (2s_{34}) \right) \right]^n \, .
\end{equation}
Formally, we expect that the sum over all orders in the derivative expansion should capture information about the UV physics
\begin{align}
   A^+(s_1\dots s_4) &= \sum_{n=0}^{\infty} A^+_n(s_1 \dots s_4) \\
   &= \sum_{n=0}^{\infty} 2\deltabr(-3-s_T) \frac{i\lambda^2}{M^2} \left[ -\frac{H^2}{M^2} \left( (2s_{34})^2 - 3 (2s_{34}) \right) \right]^n\,,
\end{align}
and although the series of correlators at the end of inflation corresponding to the right-hand sum is asymptotic, as discussed in the previous section, this series in Mellin space converges absolutely as long as $\abs{s_{34}^2 - 3s_{34} } < M^2/H^2$.
We can then analytically continue the sum to
\begin{equation}
   A^+_\text{resum} = 2\deltabr(-3-s_T) \frac{i\lambda^2}{M^2 + H^2\left( (2s_{34})^2 - 3(2s_{34}) \right)}\,,
\end{equation}
and similarly
\begin{equation}
   A^-_\text{resum} = 2\deltabr(-3-s_T) \frac{-i\lambda^2}{M^2 + H^2\left( (2s_{34})^2 - 3(2s_{34}) \right)}\,.
\end{equation}
To find the trispectrum at the end of inflation, we just need to introduce the external propagators, whereupon we recover the integral~\eqref{eq:B4s EFT integral} from the main text.

\newpage

\bibliographystyle{JHEP}
\bibliography{refs}

\end{document}